\documentclass[sigconf]{acmart}

\usepackage{graphicx}
\usepackage{multirow}
\usepackage{makecell}
\usepackage{float}

\usepackage{tikz}
\usepackage{amsmath}
\usepackage{amssymb}
\usepackage{enumitem}
\usepackage{subfig}

\AtBeginDocument{%
  \providecommand\BibTeX{{%
    \normalfont B\kern-0.5em{\scshape i\kern-0.1em b}\kern-0.1em\TeX}}}

\begin{document}

%%
%% The "title" command has an optional parameter,
%% allowing the author to define a "short title" to be used in page headers.
\title{Improved Code Summarization via a Graph Neural Network}

%%
%% The "author" command and its associated commands are used to define
%% the authors and their affiliations.
%% Of note is the shared affiliation of the first two authors, and the
%% "authornote" and "authornotemark" commands
%% used to denote shared contribution to the research.
%\author{Sakib Haque, Alex LeClair, Aakash Bansal, and Collin McMillan}
%\email{{one, two, three, cmc}@nd.edu}
%\affiliation{%
%  \institution{Department of Computer Science\\University of Notre Dame}
%  \city{Notre Dame}
%  \state{IN}
%  \country{USA}
%  \postcode{46656}
%}
\author{Alexander LeClair}
\email{aleclair@nd.edu}
\affiliation{
    \institution{University of Notre Dame}
    \city{South Bend}
    \state{IN}
    \postcode{46556}
}
\author{Sakib Haque}
\email{shaque@nd.edu}
\affiliation{
    \institution{University of Notre Dame}
    \city{South Bend}
    \state{IN}
    \postcode{46556}
}
\author{Lingfei Wu}
\email{wuli@us.ibm.com}
\affiliation{
    \institution{IBM Research}
    \city{Yorktown Heights}
    \state{NY}
    \postcode{10598}
}
\author{Collin McMillan}
\email{cmc@nd.edu}
\affiliation{
    \institution{University of Notre Dame}
    \city{South Bend}
    \state{IN}
    \postcode{46556}
}

%%
%% By default, the full list of authors will be used in the page
%% headers. Often, this list is too long, and will overlap
%% other information printed in the page headers. This command allows
%% the author to define a more concise list
%% of authors' names for this purpose.
\renewcommand{\shortauthors}{LeClair, et al.}

%%
%% The abstract is a short summary of the work to be presented in the
%% article.
\begin{abstract}
Automatic source code summarization is the task of generating natural language descriptions for source code. Automatic code summarization is a rapidly expanding research area, especially as the community has taken greater advantage of advances in neural network and AI technologies. In general, source code summarization techniques use the source code as input and outputs a natural language description. Yet a strong consensus is developing that using structural information as input leads to improved performance. The first approaches to use structural information flattened the AST into a sequence. Recently, more complex approaches based on random AST paths or graph neural networks have improved on the models using flattened ASTs. However, the literature still does not describe the using a graph neural network together with source code sequence as separate inputs to a model. Therefore, in this paper, we present an approach that uses a graph-based neural architecture that better matches the default structure of the AST to generate these summaries. We evaluate our technique using a data set of 2.1 million Java method-comment pairs and show improvement over four baseline techniques, two from the software engineering literature, and two from machine learning literature.
\end{abstract}

%%
%% The code below is generated by the tool at http://dl.acm.org/ccs.cfm.
%% Please copy and paste the code instead of the example below.
%%

% \begin{CCSXML}
% 	<ccs2012>
% 	<concept>
% 	<concept_id>10011007</concept_id>
% 	<concept_desc>Software and its engineering</concept_desc>
% 	<concept_significance>500</concept_significance>
% 	</concept>
% 	<concept>
% 	<concept_id>10011007.10011006.10011073</concept_id>
% 	<concept_desc>Software and its engineering~Software maintenance tools</concept_desc>
% 	<concept_significance>500</concept_significance>
% 	</concept>
% 	</ccs2012>
% \end{CCSXML}

% \ccsdesc[500]{Software and its engineering}
% \ccsdesc[500]{Software and its engineering~Software maintenance tools}

%%
%% Keywords. The author(s) should pick words that accurately describe
%% the work being presented. Separate the keywords with commas.

%%
%% This command processes the author and affiliation and title
%% information and builds the first part of the formatted document.
\maketitle

\vspace{-0.2cm}
\section{Introduction}

Source code summarization is the task of writing brief natural language descriptions of code~\cite{leclair2019neural, haiduc2010use, mcburney2016automatic, eddy2013evaluating}.  These descriptions have long been the backbone of developer documentation such as JavaDocs~\cite{kramer1999api}.  The idea is that a short description allows a programmer to understand what a section of code does and that code's purpose in the overall program, without requiring the programmer to read the code itself.  Summaries like ``uploads log files to the backup server'' or ``formats decimal values as scientific notation'' can give programmers a clear picture of what code does, saving them time from comprehending the details of that code.

Automatic code summarization is a rapidly expanding research area.  Programmers are notorious for neglecting the manual effort of writing summaries themselves~\cite{deSouza:2005:SDE:1085313.1085331, Shi:2011:ESE:1987434.1987473, Kajko-Mattsson:2005:SDP:1032622.1035374, Roehm:2012:PDC:2337223.2337254}, and automation has long been cited as a desirable alternative~\cite{Forward:2002:RSD:585058.585065}.  The term ``source code summarization'' was coined around ten years ago~\cite{haiduc2010use} and since that time the field has proliferated.  At first, the dominant strategy was based on sentence templates and heuristics derived from empirical studies~\cite{mcburney2016automated, sridhara2010towards, sridhara2011automatically, rodeghero2015eye, moreno2013automatic, eddy2013evaluating}.  Starting around 2016, data-driven strategies based on neural networks came to the forefront, leveraging gains from both the AI/NLP and mining software repositories research communities~\cite{iyer2016summarizing, hu2018deep, leclair2019neural, alon2018code2seq}.

These data-driven approaches were inspired by neural machine translation (NMT) from natural language processing.  In NMT, a sentence in one language e.g. English is translated into another language e.g. Spanish.  A dataset of millions of examples of English sentences paired with Spanish translations is required.  A neural architecture based on the encoder-decoder model is used to learn the mapping between words and even the correct grammatical structure from one language to the other based on these examples.  This works well because both input and output languages are sequences of roughly equal length, and mappings of words tend to exist across languages.  The metaphor in code summarization is to treat source code as one language input and summaries as another.  So code would be input to the same models' encoder, and summaries to the decoder.  Advances in repository mining made it possible to gather large datasets of paired examples~\cite{Lopes+Bajracharya+Ossher+Baldi:2010}.

But evidence is accumulating that the metaphor to NMT has major limits~\cite{hellendoorn2017deep}.  Source code has far fewer words that map directly to summaries than the NMT use case~\cite{leclair2019recommendations}.  Source code tends not to be of equal length to summaries; it is much longer~\cite{moreno2012analysis, eddy2013evaluating}.  And crucially, source code is not merely a sequence of words.  Code is a complex web of interacting components, with different classes, routines, statements, and identifiers connected via different relationships.  Software engineering researchers have long recognized that code is much more suited to graph or tree representations that tease out the nuances of these relationships~\cite{binkley2007source, ottenstein1984program}.  Yet, the typical application of NMT to code summarization treats code as a sequence to be fed into a recurrent neural network (RNN) or similar structure designed for sequential information.

The literature is beginning to recognize the limits to sequential representations of code for code summarization.  Hu~\emph{et al.}~\cite{hu2018deep} annotate the sequence with clues from the abstract syntax tree (AST).  LeClair~\emph{et al.}~\cite{leclair2019neural} expand on this idea by separating the code and AST into two different inputs.  Alon~\emph{et al.}~\cite{alon2018code2seq} extract paths from the AST to aid summarization.  Meanwhile, Allamanis~\emph{et al.}~\cite{allamanis2018learning} propose using graph neural networks (GNNs) to learn representations of code (though for the problem of code generation, not summarization).  These approaches all show how neural networks can be effective in extracting information from source code better in a graph or tree form than in a sequence of tokens, and using that information for downstream tasks such as summarization.

What is missing from the literature is a thorough examination of \textbf{how} graph neural networks improve representations of code based on the AST.  There is evidence \emph{that} GNN-based representations improve performance, but the degree of that improvement for code summarization has not been explored thoroughly, and the reasons for the improvement are not well understood.

In this paper, we present an approach for improving source code summarization using GNNs.  Specifically, we target the problem of summarizing program subroutines.  Our approach is based on the {\small \texttt{graph2seq}} model presented by Xu~\emph{et al.}~\cite{xu2018graph2seq}, though with a few modifications to customize the model to a software engineering context.  In short, we use the GNN-based encoder of {\small \texttt{graph2seq}} to model the AST of each subroutine, combined with the RNN-based encoder used by LeClair~\emph{et al.}~\cite{leclair2019neural} to model the subroutine as a sequence.  We demonstrate a 4.6\% BLEU score improvement for a large, published dataset~\cite{leclair2019recommendations} as compared to recent baselines.  In an experiment, we use techniques from the literature on explainable AI to propose explanations for why the approach performs better and in which cases.  We seek to provide insights to guide future researchers.  We make all our data, implementations, and experimental framework available via our online appendix (Section~\ref{sec:repro}).
\vspace{-0.5cm}
\section{Problem, Significance, Scope}

We target the problem of automatically generating summaries of program subroutines. To be clear, the input is the source code of a subroutine, and the output is a short, natural language description of that subroutine. These summaries have several advantages when put into documentation such as decreased time to understand code \cite{Forward:2002:RSD:585058.585065}, improved code comprehension \cite{von1995program,cornelissen2009systematic}, and to making code more searchable \cite{howard2013mappings}. Programmers are notorious for consuming high quality documentation for themselves, while neglecting to write and update it themselves \cite{Forward:2002:RSD:585058.585065}. Therefore, recent research has focused on automating the documentation process. Current research has had success generating summaries for a subset of methods that are generally shorter and use simpler language in both the code and reference comment (e.g. setters and getters), but have had a problem with methods that have more complex structures or language \cite{leclair2019neural}. A similar situation for other SE research problems has been helped by various graph representations of code \cite{fernandes2019structured, allamanis2018learning}, but using graph representations is only starting to be accepted for code summarization \cite{alon2018code2seq, fernandes2019structured}.
Graph representations have the potential to improve code summarization because, instead of using only a sequence of code tokens as input, the model can access a rich variety of relationships among tokens. 

Automatic documentation has a large potential impact on how software is developed and maintained. Not only would automatic documentation reduce the time and energy programmers spend reading and writing software, having a high level summary available has been shown to improve results in other SE tasks such as code categorization and code search \cite{leclair2018codecat, howard2013mappings}.

\vspace{-0.2cm}
\section{Background and Related Work}
\label{sec:background}

This section discusses some of the previous work relevant to this work and source code summarization. 

\vspace{-0.2cm}
\subsection{Source Code Summarization}
\label{sec:scsbackground}

Source code summarization research can be broadly categorized as either 1) heuristic/template-driven approaches or 2) more recent AI/Data-driven approaches. Heuristic-based approaches for source code summarization started to gain popularity in 2010 with work done by Haiduc \emph{et al.} \cite{Haiduc:S:ICPC:2008}. In their work, text retrieval techniques and latent semantic indexing (LSI) were used to pick important keywords out of source code, then those words are considered the summary. Early work done by Haiduc \emph{et al.} and others have helped inspire other work using extractive summarization techniques based on TF-IDF, LSI, and LDA to create a summary. Heuristic-based approaches are less related to this work than data-driven approaches, so due to space limitations we direct readers to surveys by Song \emph{et al.} \cite{song2019survey} and Nazar \emph{et al.} \cite{nazar2016summarizing} for additional background on the topic.

This paper builds on the current work done with data-driven approaches in source code summarization which have dominated NLP and SE literature since around 2015.  In Table \ref{tab:scs_compare}, we divide recent work into two groups by their use of the AST as an input to the model. Then we further divide related work by the following six attributes:

\begin{enumerate}[leftmargin=*]
    \item Src Code - A model uses the source code sequence as input, not as part of the AST.
    \item AST - A model uses the AST as input.
    \item API - A model uses API information.
    \item FlatAST - Using a flattened version of the AST as model input.
    \item GNN - The model uses a form of graph neural network for node/edge embedding.
    \item Paths - Using a path through the AST as input to the model.
    
\end{enumerate}

A brief history of the related data-driven work starts with Iyer \emph{et al} \cite{iyer2016summarizing}. In their work they used stack overflow questions and responses where the title of the post was considered the high level summary, and the source code in the top rated response was used as the input. The model they developed was an attention based sequence to sequence model similar to those used in neural machine translation tasks \cite{sutskever2014sequence}. To expand on this idea, Hu \emph{et al.} \cite{hu2018summarizing} added API information as an additional input into the model. They found that the model was able to generate better responses if it had access to information provided by API calls in the source code.

Later, Hu \emph{et al.} \cite{hu2018deep} developed a structure based traversal (SBT) method for flattening the AST into a sequence that keeps words in the code associated with their node type. The SBT sequence is a combination of source code tokens and AST structure which was then input into an off the shelf encoder/decoder model. LeClair \emph{et al.} \cite{leclair2019neural} built upon this work by creating a multi-input model that used the SBT sequence with all identifiers removed as the first input, and the source code tokens as the second. They found that if you decouple the structure of the code form the code itself that the model improved its ability to learn that structure.

More recently, Alon \emph{et al.} \cite{alon2018code2seq} in 2018 proposed a source code summarization technique that would encode each pairwise path between nodes in the AST. They would then randomly select a subset of these paths for each iteration in training. These paths were then encoded and used as input to a standard encoder/decoder model. They found that encoding the AST paths allowed the model to generalize to unseen methods more easily, as well as providing a level of regularization by randomly selecting a subset of paths each training iteration.

 Then in 2019 Fernandes \emph{et al.} \cite{fernandes2019structured} developed a GNN based model that uses three graph representations of source code as input 1) next token, 2) AST, and 3) last lexical use. To represent these three graphs they used a shared node setup where each graph represented a different set of edges between source code tokens. Using this approach they observed a better ``global'' view of the method and had success with maintaining the central named entity from the method. Fernandes' observation is an important clue that there is additional information embedded in the source code beyond the sequence of tokens, this motivates the use of a GNN for the AST as a separate input in our work. 

\begin{table}[t!]
  % \centering

  \scalebox{0.82}{%
    \begin{tabular}{c|c|c|c|c|c|c}

         &Src Code&AST&API&FlatAST&GNN&Paths\\
         \hline
         2016 Iyer \emph{et al.} \cite{iyer2016summarizing}&x&&&&& \\
         2017 Loyola \emph{et al.} \cite{Loyola2017ANA}&x&&&&&\\
         2017 Lu \emph{et al.} \cite{lu2017learning}&x&&x&&&\\
         2018 Hu \emph{et al.} \cite{hu2018summarizing}&x&&x&&& \\
        2018 Liang \emph{et al.} \cite{liang2018automatic}&x&x&&&&\\
        2018 Hu~\emph{et al.} \cite{hu2018deep}&x&x&&x&&\\
        2018 Wan~\emph{et al.} \cite{wan2018reinforcement}&&x&&&x&\\
        2019 LeClair~\emph{et al.} \cite{leclair2019neural}&x&x&&x&&\\
        2019 Alon~\emph{et al.} \cite{alon2018code2seq}&&x&&x&&x\\
        2019 Fernandes~\emph{et al.} \cite{fernandes2019structured}&&x&&&x&\\
        % \hline
    \end{tabular}
    }
    \vspace{0.1cm}
    \caption{Comparison of recent data-driven Source Code Summarization research categorized by the data, architectures, and approaches used. The approaches in the upper table use only the source code sequence as input to their models, while the bottom table approaches use the AST or a combination of AST and source code.}
    \label{tab:scs_compare}
    \vspace{-0.8cm}
\end{table}

\vspace{-0.1cm}
\subsection{Neural Machine Translation}

For the last six years work in neural machine translation (NMT) has been dominated by the encoder-decoder model architecture developed by Bahdanau \emph{et al.} \cite{bahdanau2014neural}. The encoder-decoder architecture can be thought of as two separate models, one to encode the input (e.g. English words) into a vector representation, and one to decode that representation into the desired output (e.g. German tokens). 

Commonly, encoder-decoder models use a recurrent layer (RNN, GRU, LSTM, etc.) in both the encoder and decoder. Recurrent layers are effective at learning sequence information because for each token in a sequence, information is propagated through the layer \cite{sutskever2011generating}. This allows each token to affect the following tokens in the sequence. Some of the common recurrent layers such as the GRU and LSTM also can return state information at each time step of the input sequence. The state information output from the encoder is commonly used to seed the initial state of the decoder improving translation results \cite{sutskever2014sequence}.

Another more recent addition to many encoder-decoder models is the attention mechanism. The intuition behind attention is that not all tokens in a sequence are of equal importance to the final output prediction. What attention tries to do is to learn what words are important and map input tokens to output tokens. It does this by taking the input sequence at every time step and the predicted sequence at a time step and tries to determine which time step in the input will be most useful to predict the next token in the output.

\begin{figure}[b!]
	\vspace{-0.1cm}
	\centering
	\includegraphics[width=0.96\columnwidth]{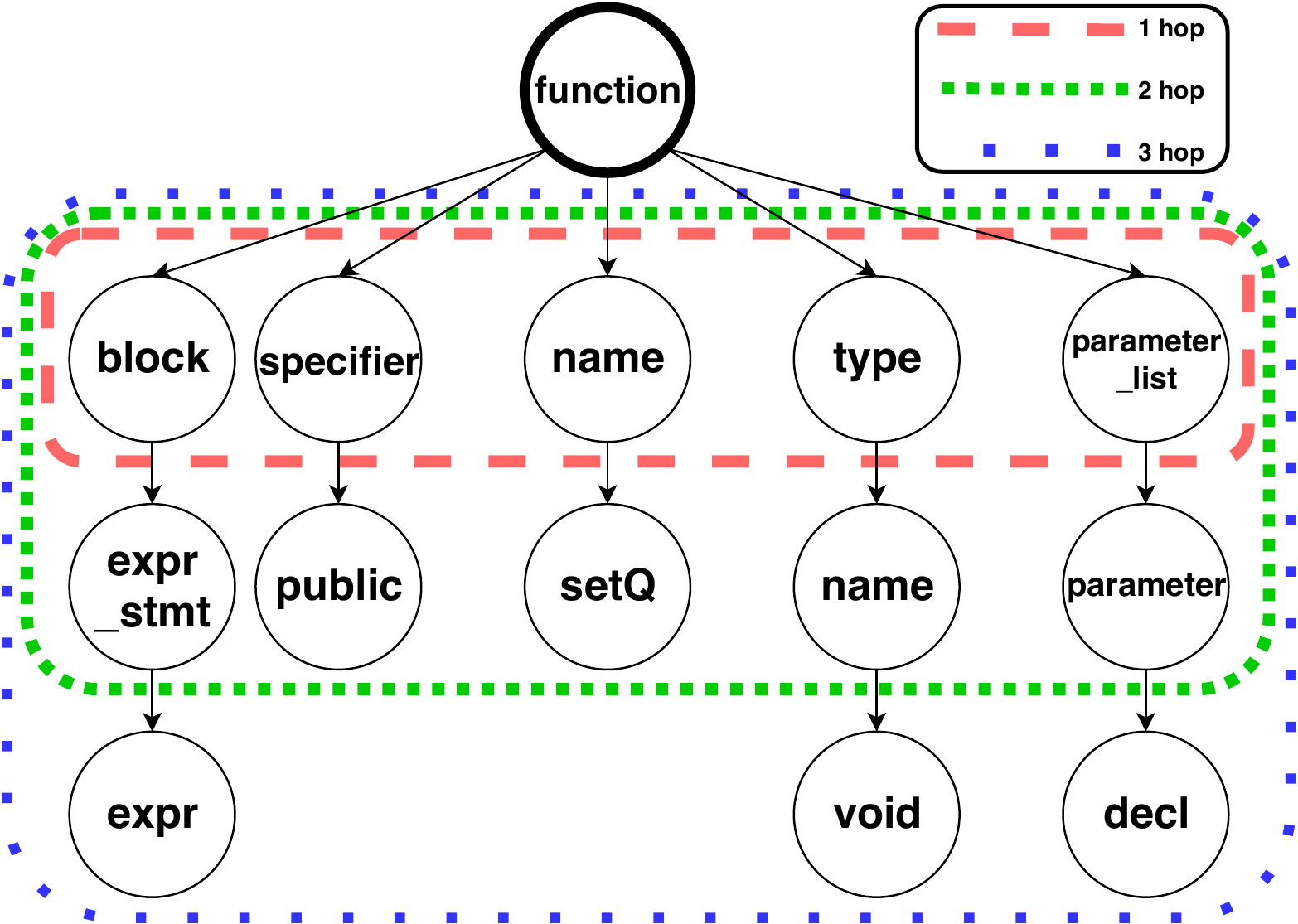}
	\setcounter{figure}{0}
	\vspace{-0.3cm}
	\label{fig:asthops}
\end{figure}

\begin{figure*}[!t]
	\vspace{-0.1cm}
	\centering
	\includegraphics[width=0.8\textwidth]{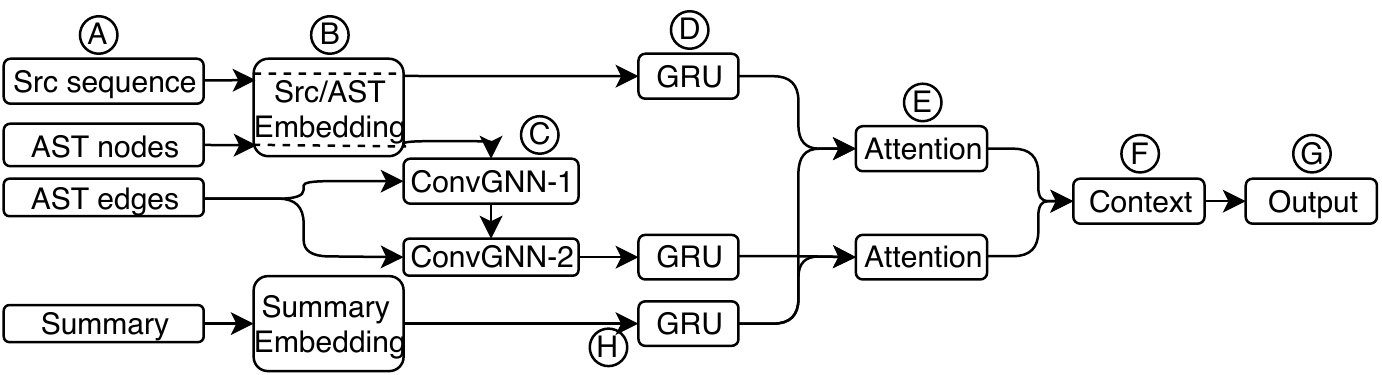}
	\vspace{-0.1cm}
	\caption{High level diagram of model architecture for 2-hop model}
	\label{fig:model}
	\vspace{-0.2cm}
\end{figure*}

\vspace{-0.1cm}
\subsection{Graph Neural Networks}
\label{sec:gnn}
Graph Neural Networks are another key background technology to this paper.  A recent survey by Wu et al. \cite{wu2019gnnsurvey} categorizes GNNs into four groups:

\vspace{-0.1cm}
\begin{enumerate}
    \item Recurrent Graph Neural Networks (RecGNNs)
    \item Convolutional Graph Neural Networks (ConvGNNs)
    \item Graph Autoencoders (GAEs)
    \item Spatial-temporal Graph Neural Networks (STGNNs)
\end{enumerate}
\vspace{-0.1cm}

We will focus on ConvGNNs in this section because they are well suited for this task, and it is what we use in this paper. ConvGNNs were developed after RecGNNs and were designed with the same idea of message passing between nodes. They have also been shown to encode spatial information better than RecGNNs and are able to be stacked, improving the ability to propagate information across nodes \cite{wu2019gnnsurvey}. ConvGNNs take graph data and learn representations of nodes based on the initial node vector and its neighbors in the graph. The process of combining the information from neighboring nodes is called ``aggregation.'' By aggregating information from neighboring nodes a model can learn representations based on arbitrary relationships. These relationships could be the hidden structures of a sentence, the parts of speech \cite{chen2019reinforcement}, dependency parsing trees \cite{xu2018exploiting}, or the sequence of tokens \cite{chen2017improved}. ConvGNNs have been used for similar tasks before, such as in graph2seq for semantic parsing \cite{xu2018graph2seq} and natural question generation \cite{chen2019reinforcement}.

ConvGNNs also allow nodes to get information from other nodes that are further than just a single edge or ``hop'' away. In the figure below we show an example partial AST and what 1, 2, and 3 hops look like for the token `function'. Each time a hop is performed, the node gets information from its neighboring nodes. So, on the first hop the token `function' aggregates information from the nodes `block', `specifier', `name', `type', and `parameter\_list'. In the next hop that occurs, the `function' node will still only combine information from its neighbors, but now each of those nodes will have aggregate information from their children. For example, the node `block' will contain information from the `expr\_stmt' node. Then when the `function' node aggregates the `block' node, it has information from both `block' and `expr\_stmt'.

There are several aggregation strategies which have been shown to have different performance for different tasks \cite{wu2019gnnsurvey, fernandes2019structured}. A common aggregation strategy is to sum a node vector with its neighbors and then apply an activation on that node, but there are many schemes that can be used to combine node information. Some other approaches to this are pooling, min, max, and mean. Xu~\emph{et al.} discuss different node and edge aggregation methods in their paper on creating sequences from graphs. They found that in most cases a mean aggregator out performed other types of aggregators, including one using an LSTM.

\vspace{-0.2cm}
\section{approach}
\label{sec:approach}

This section provides the details of our approach. Our model is based off the neural model proposed by LeClair \emph{et al.} \cite{leclair2019neural} and builds on that work by using ConvGNNs discussed in the previous section. In a nutshell, our approach works in 5 steps: 
\vspace{-0.1cm}
\begin{enumerate}
    \item Embed the source code sequence and the AST node tokens.
    \item Encode the embedding output with a recurrent layer for the source code token sequence and a ConvGNN for the AST nodes and edges.
    \item Use an attention mechanism to learn important tokens in the source code and AST.
    \item Decode the encoder outputs.
    \item Predict the next token in the sequence.
\end{enumerate}
\vspace{-0.1cm}

\vspace{-0.1cm}
\subsection{Model Overview}

An overview of our model is in Figure~\ref{fig:model}. In a nutshell, what we did was modify the model on the multi-input encoder-decoder proposed by LeClair~\emph{et al.} to use a ConvGNN instead of a flattened AST. Notice in area A of Figure~\ref{fig:model} that our model has four inputs 1) the sequence of source code tokens, 2) the nodes of the AST, 3) the edges of the AST, 4) the predicted sequence up to this point. Next, in area B, we embed the inputs using standard embedding layers. The source sequence and AST nodes share an embedding due to a large overlap in vocabulary. Then in area C of Figure~\ref{fig:model} the AST nodes are input into the ConvGNN layers, the number of layers here depends on the hop size of the model, and then input into a GRU. The source code sequence goes into a GRU after the embedding in area D. For the decoder in area H, we have an embedding layer feeding into a GRU. We then do two attention mechanisms seen in area E, one between the source code and summary, and the other between the AST and summary. Then in areas F and G we combine the outputs of our attention creating a context vector which is flattened and used to predict the next token in the sequence.
\vspace{-0.2cm}
\paragraph{Key Novel Component} The key novel component of this paper is in processing the AST using a ConvGNN and combining the output of the ConvGNN encoder with the output of the source code token encoder. In our approach, the ConvGNN allows the nodes of the AST to learn representations based on their neighboring nodes. Teaching the model information about the structure of the code, and how it relates to the tokens found in the source code sequence. Both the source and and AST encodings are input into separate attention mechanisms with the decoder and are then concatenated. This creates a context vector which we then use in a dense layer to predict the next token in the sequence.

Basically, what we do is combine the structure of the sequence (the AST) with the sequence itself (the source code). Combining the structure of a sequence and the sequence itself into a model has been shown to improve the quality of generated summaries in both SE and NLP literature \cite{hu2018deep, leclair2019neural}. In this paper we aim to show that using a neural network architecture that is more suited to the structure of the data (graph vs sequence) we can further improve the models ability to learn complex relationships in the source code.

\vspace{-0.2cm}
\subsection{Model Details}

In this section we will discuss specific model implementation details that we used for our best performing model to encourage reproducibility. We developed our proposed model using Keras~\cite{chollet2015keras} and Tensorflow~\cite{tensorflow2015whitepaper}. We also provide our source code and data as an online appendix (details can be found in Section~\ref{sec:repro}). 

First, as mentioned in the previous section, our model is based on the encoder-decoder architecture and has four inputs 1) the source code sequence, 2/3) the AST as a collection of nodes along with an adjacency matrix with edge information and 4) the comment generated up to this point which is the input to the decoder. As seen in Figure~\ref{fig:model}, we use two embedding layers one for the source code and AST and one for the decoder. We use a single embedding layer for both the source code and AST node inputs because they have such a large overlap in vocabulary. The shared embedding layer has a vocabulary size of 10908 and an embedding size of 100. The decoder embedding layer has a vocabulary size of 10000 and an embedding size of 100. So far, this follows the model proposed by LeClair \emph{et al.} \cite{leclair2019neural}.

Next, the model has two encoders, one for the source code sequence and another for the AST. The source code sequence encoder is a single GRU layer with an output length of 256. We have the source code GRU return its hidden states to use as the initial state for the decoder GRU. The second encoder, the AST encoder, is a collection of ConvGNN layers followed by a GRU of length 256. The number of ConvGNN layers depends on the number of hops used, for our best model this was 2-hops as seen in Figure~\ref{fig:model}.

The ConvGNN that we use for the AST node embeddings takes the AST embedding layer output and the AST edge data as inputs. Then, for each node in the input it sums the current node vector with each of it's neighbors and multiplies that by a set of trainable weights and adds a trainable bias. In our best performing implementation we use a ConvGNN layer for each hop in the model as seen in Figure~\ref{fig:model}. We also test our model with different numbers of hops, which slightly changes the architecture of the AST encoder of the model by adding additional ConvGNN layers.

Next, we have two attention mechanisms 1) an attention between the decoder and the source code, and 2) between the decoder and the AST. These attention mechanisms learn which tokens from the source code/AST are important to the prediction of the next token in the decoder sequence given the current predicted sequence generated up to this point. The attention mechanisms are then concatenated together with the decoder to create a final context vector. Then, we apply a dense layer to each vector in our final context vector, which we then flatten and use to predict the next token in the sequence.

\vspace{-0.2cm}
\subsection{Data Preparation}

The data set that we used for this project was provided by LeClair \emph{et al.} in a paper on recommendations for source code summarization datasets \cite{leclair2019recommendations}. LeClair \emph{et al.} describe best practices for developing a dataset for source code summarization and also provide a dataset of 2.1 million Java method comment pairs. They provide their dataset in two versions, 1) a filtered version with the raw, unprocessed version of the methods and comments and 2) the tokenized version where text processing has already been applied. For our baseline comparisons, we use the tokenized version of the dataset provided by LeClair \emph{et al.} allowing us to directly compare results with their work in source code summarization. The dataset did not include ASTs that were already parsed, so we use the SrcML library \cite{collard2011lightweight} to generate the associated ASTs from the raw source code.

\vspace{-0.2cm}
\subsection{Hardware Details}

For training, validating, testing of our models we used a workstation with Xeon E1430v4 CPUs, 110GB RAM, a Titan RTX GPU, and a Quadro P5000 GPU. Software used include the following:

\begin{table}[!h]
\vspace{-0.3cm}
\begin{tabular}{c|c|c}
    Ubuntu 18.04 &Python 3.6& CUDA 10 \\
    Tensorflow 1.14 & Keras 2.2&CuDNN 7\\
\end{tabular}
\vspace{-0.3cm}
\end{table}

\vspace{-0.2cm}
\section{Experiment Design}

In this section we discuss the design of our experiments and discuss the methodology, baselines, and metrics used to obtain our results.

\begin{table*}[!t]
\vspace{-0.2cm}
    \centering
    \begin{tabular}{|lc|c c c c c|c|}
    \hline
    Baseline&&BLEU-A&BLEU-1&BLEU-2&BLEU-3&BLEU-4&ROUGE-LCS F1 \\
    \hline
        ast-attendgru && 18.69 & 37.13 & 21.11 & 14.27 & 10.90 & 49.75 \\
        graph2seq &&18.61&37.56&21.27&14.13&10.63&49.69\\
        code2seq && 18.84 & 37.49& 21.36 & 14.37 & 10.95 & 49.69 \\
        BiLSTM+GNN->LSTM && 19.05 & 37.70 & 21.53 & 14.59 & 11.11 & 55.74 \\
        \hline
        ConvGNN Models&\# of hops&BLEU-A&BLEU-1&BLEU-2&BLEU-3&BLEU-4&ROUGE-LCS F1\\
        \hline
        code+gnn+dense &2 & 19.46 & 38.71 & 22.04 & 14.86 & 11.31 & 56.07 \\
        code+gnn+BiLSTM & 2& \textbf{19.93} & \textbf{39.14} & \textbf{22.49} & \textbf{15.31} & 11.70 & 56.08 \\
        code+gnn+GRU & 1 & 19.70 & 38.15 & 22.12 & 15.22 & \textbf{11.73} & \textbf{57.15} \\
        code+gnn+GRU & 2 & 19.89 & 39.01 & 22.42 & 15.28 & 11.70 & 55.78 \\
        code+gnn+GRU & 3 & 19.58 & 38.48 &22.09 & 15.01 & 11.52 & 56.14 \\
        code+gnn+GRU & 5 & 19.68 & 38.89 & 22.30 & 15.09 & 11.46 & 55.81 \\
        code+gnn+GRU & 10 & 19.34 & 38.68 &21.94 & 14.73 & 11.20 & 55.10 \\
        \hline
         
    \end{tabular}
    \vspace{0.2cm}
    \caption{BLEU and ROUGE-LCS scores for the baselines and our proposed models}
    \label{tab:metrics}
    \vspace{-0.8cm}
\end{table*}

\vspace{-0.1cm}
\subsection{Research Questions}
\label{sec:rqs}

Our research objective is to determine if our proposed approach of using the source code sequence along with a graph based AST and ConvGNN outperform current baselines. We also want to determine why our proposed model may outperform current baselines based on the use of the AST graph and ConvGNN. We ask the following Research Questions (RQs) to explore these situations:
\vspace{-0.1cm}
\begin{description}
	\item[RQ$_{1}$] What is the performance of our approach compared to the baselines in Section~\ref{sec:baselines} in terms of the metrics in Section~\ref{sec:metrics}?
	
	\vspace{0.05cm}
	
	\item[RQ$_{2}$] What is the degree of difference in performance caused by the number of graph hops in terms of the metrics in Section~\ref{sec:metrics}?
	
	\vspace{0.05cm}
	
	\item[RQ$_{3}$] Is there evidence that the performance differences are due to use of the ConvGNN?
\end{description}
\vspace{-0.1cm}
The rationale for RQ$_1$ is to compare our approach with other approaches that use the AST as input. Previous work has already shown that the inclusion of the AST as an input to the model outperforms previous models where no AST information was provided \cite{hu2018summarizing, leclair2019neural, alon2018code2seq, fernandes2019structured}. Some previous work provides the AST as a tree or graph \cite{alon2018code2seq, fernandes2019structured}, but the source code sequence was not provided to the model. Our proposed model is a logical next step in source code summarization literature and we ask RQ$_1$ to evaluate our model against previous models.

The rationale for RQ$_2$ is to determine what affect (if any) the number of hops has on the generated summaries (a description of ConvGNN hops can be found in Section~\ref{sec:gnn}). Xu~\emph{et al.} \cite{xu2018graph2seq} discuss the impact of hop size on their work with generating SQL queries. They found that for their test models, the number of hops did not affect model convergence. To test this they generate random directed graphs of sizes 100 and 1000 and trained a model to find the shortest path between nodes, but did not evaluate how hop size affects the task of source code summarization. Since ConvGNNs create a layer for each hop, it becomes computationally expensive to train models with an arbitrarily large number of hops. With RQ$_2$ we hope to build an intuition into how the number of hops affects ConvGNN learning specifically for source code summarization. 

The rationale for RQ$_3$ is that discovering \emph{why} a model learned to generate certain summaries can be just as important as evaluation metrics \cite{samek2017explainable,doshi2017interpretableml,MILLER2019explainai,doran2017explainable,rueden2019informed, Arras2017explainabletext}. Doshie~\emph{et al.} \cite{doshi2017interpretableml} discuss what interpretability means and offers guidelines to researchers on what they can do to make their models more explainable, while Roscher~\emph{et al.} state that \emph{``...explainability is a prerequisite to ensure the scientific value of the outcome''}\cite{roscher2019explainable}. As models are developed many factors change, and it is often times not an easy task to determine which factors had the greatest impact on performance. In their work on explainable AI, Arras~\emph{et al.} and Samek~\emph{et al.} show how you can use visualizations to aid in the process of explainability for text based modeling tasks \cite{Arras2017explainabletext, samek2017explainable}. With RQ$_3$ we aim to explain what impact the inclusion of the AST as a graph and ConvGNN had on generated summaries. 

\vspace{-0.2cm}
\subsection{Methodology}

To answer RQ$_1$, we follow established methodology and evaluation metrics that have become standard in both source code summarization work  and neural machine translation from NLP \cite{papineni2002bleu,lin2004rouge}. To start, we use a large well-documented data set from the literature to allow us to easily compare results and baselines. We use the data handling guidelines outlined in LeClair \emph{et al.} \cite{leclair2019recommendations} so that we do not have data leakage between our training, validation, and testing sets. Next, we train our models for ten epochs and choose the model with the highest validation accuracy score for our comparisons. Choosing the model with the best validation performance out of ten epochs is a training strategy that has been successfully used in other related work \cite{leclair2019neural}. For RQ$_1$ we evaluate the best performing model using automated evaluation techniques to compare against our baselines and report in this paper. 

For RQ$_2$ we train five ConvGNN models with all hyper-parameters frozen except for hop size. We test our model using hop sizes of 1,2,3,5, and 10 in line with other related work \cite{xu2018graph2seq}. We used the model configuration outlined in Section~\ref{sec:approach} that uses the source code and AST input, as well as a GRU layer directly after the ConvGNN layers. We chose this model because of its performance and its faster training speed compared with the BiLSTM model. To report our results we use the same ``best of ten'' technique that we use to answer RQ$_1$, that is, we train each model for ten epochs and report the results on the model with the highest validation accuracy.

For RQ$_3$ we use a combination of automated tools and metrics such as BLEU~\cite{papineni2002bleu}, ROUGE~\cite{lin2004rouge}, and visualizations from model weights. Visualizing model weights and parameters has become a popular way to help explain what deep learning models are doing, and possibly give insight into why they generate the output that they do. To help us answer RQ$_3$ we use concepts similar to those outlined in Samek~\emph{et al.} \cite{samek2017explainable} for model visualizations.

\vspace{-0.6cm}
\subsection{Metrics}
\label{sec:metrics}

For our quantitative metrics we use both BLEU \cite{papineni2002bleu} and ROUGE \cite{lin2004rouge} to evaluate our model performance. BLEU scores are a standard evaluation metric in the source code summarization literature \cite{leclair2019neural, hu2018summarizing, iyer2016summarizing}. BLEU is a text similarity metric that compares overlapping n-grams between two given texts. While BLEU can be thought of as a precision score: how much of the generated text appears in the reference text. In contrast, ROUGE can be thought of as a recall score: how much of the reference appears in the generated text. 

ROUGE is used primarily in text summarization tasks in the NLP literature due to the score allowing multiple references since there may be multiple correct summaries of a text \cite{lin2004rouge}. In our work we do not have multiple reference texts per method, but ROUGE gives us additional information about the performance of our models that BLEU scores alone do not provide. In this paper we report a composite BLEU score, BLEU$_1$ through BLEU$_4$ (n-grams of length 1 to length 4), and ROUGE-LCS (longest common sub-sequence) to have a well rounded set of automated evaluation metrics.

\vspace{-0.3cm}
\subsection{Baselines}
\label{sec:baselines}

We compare our model against four baselines. These baselines are all from recent work that is directly relevant to this paper. We chose these baselines because they provide comparison for three categories in source code summarization using the AST: 1) flattened AST, 2) using paths through the AST, and 3) using a graph neural network to encode the AST.

Each of these baselines uses AST information as input to the model with different schemes. They also cover a variety of model architectures and configurations. Due to space limitations we do not list all relevant details for each model, but have a more in depth overview in Section \ref{sec:scsbackground} and in Table \ref{tab:scs_compare}.

\vspace{-0.1cm}
\begin{itemize}
 \item \textbf{ast-attendgru}: In this model LeClair \emph{et al.} \cite{leclair2019neural} use a standard encoder-decoder model and add an additional encoder for the AST. They flatten the AST using the SBT technique outline in Hu \emph{et al.} \cite{hu2018deep}. Then both the source code tokens and the flattened AST are provided as input into the model. For encoding these inputs they use recurrent layers and then use a decoder with a recurrent layer to generate the predictions. This approach is representative of other approaches that flatten the AST into a sequence.

\item \textbf{graph2seq}: Xu \emph{et al.} \cite{xu2018graph2seq} developed a general graph to sequence neural model that generates both node and graph embeddings. In their work they use an SQL query and generate a natural language query based on the SQL. Their implementation propagates both forward and backwards over the graph, and includes node level attention. They achieved state of the art results on an SQL->natural language task using BLEU-4 as a metric. They also evaluate how the number of hops affected the performance of the model finding that any number of hops still converged to similar results, but specific models could perform just as well with less hops lowering the amount of computation needed.

\item \textbf{code2seq}: Alon \emph{et al} \cite{alon2018code2seq} use random pairwise paths through the AST as model input which we discuss more in depth in Section \ref{sec:scsbackground}. They used C\# code to generate summaries, while we use Java. They had a variety of configurations that they test, due to this we did a good-faith re-implementation of their base model in an attempt to capture the major contributions of their approach.

\item \textbf{BILSTM+GNN}: Fernandes \emph{et al.} \cite{fernandes2019structured} proposed a model using a BILSTM and GNN trained with a C\# data set for code summarization. We reproduced a model using the information outlined in their paper. We trained the model using the Java data set from LeClair \emph{et al.} to create a comparison for our work. In their paper they report results on a variety of model architectures and setups, we include comparison results with a model based on their best performing configuration.
\end{itemize}

These baselines are not an exhaustive list of relevant work, but they cover recent techniques used for source code summarization. Some other work that we chose not to use for baselines include Hu \emph{et al.} \cite{hu2018deep}, and Wan \emph{et al.} \cite{wan2018reinforcement}. We chose not to include Hu \emph{et al.} in our baselines because the work done by LeClair \emph{et al.} built upon their work and was shows to have higher performance, and is much closer to our proposed work in this paper. In our proposed model we use the technique outlined in LeClair \emph{et al.} of separating the source code sequence tokens from the AST.

Wan~\emph{et al.} \cite{wan2018reinforcement} is another potential baseline, but we found it unsuitable for comparison in this paper for three reasons: 1) the approach combines an AST+code encoding with Reinforcement Learning (RL), and the RL component adds many experimental variables with effects difficult to distinguish from the AST+code component, 2) the AST+code encoding technique has now been superceded by other techniques which we already use as baselines, and 3) we were unable to reproduce the results in the paper.  An interesting question for future work is to study the effects of the RL component in a separate experiment: the RL component of Wan~\emph{et al.} is supportive of, rather than a competitor with, the AST+code encoding. We also do not compare against heuristic based approaches. Most data-driven approaches outperform heuristic based approaches in all of the automated metrics, and previous work has already reported the comparisons.

\vspace{-0.3cm}
\subsection{Threats to Validity}

The primary threat to validity for this paper is that the automated metrics we use to score and rate out models may not be representative of human judgement. BLEU and ROUGE metrics can give us a good indication how our model performs compared to the reference text and other models, but there are instances where the model may generate a valid summary that does not align with the reference text. On the other hand, there is no evaluation as to whether a reference comment for a given method is a good summary. The benefit of these automated metrics is that they are fast and have wide use among the source code summarization community. To mitigate the potential pitfalls that using automated metrics may involve, we include an in depth discussion and evaluation of specific examples from our model to help interpret what our model has learned when compared to baselines. 

The dataset we use is also another potential threat to validity. While other data sets do exist with other programming languages, for example C\# or Python, many of these data sets lack the size and scope of the data set provided by LeClair \emph{et al.}. For example the C\# dataset used Fernandes \emph{et al.} has 23 projects, with 55,635 methods having associated documentation. Another common pitfall of datasets described in LeClair \emph{et al.} is that many datasets split data on the function level instead of the project level. This means that functions from the same project can appear in both the training and testing sets causing potential data leakage. Using Java is beneficial due to its widespread use in many different types of programs and its adoption in industry. Java also has a well defined commenting standard with JavaDocs that creates easily parsable documentation.

One other threat to validity is that we were unable to perform extensive hyper-parameter optimizations on our models due to hardware limitations. It could be the case that some of our models or baselines outlined in Table~\ref{tab:metrics} could be heavily impacted by certain hyper-parameters (e.g., input sequence length, learning rate, and vocabulary size), giving different scores and rankings. This is a common issue with deep learning projects, and this affects nearly all similar types of experiments. We try to mitigate the impact of this issue by being consistent with our hyper-parameter values. We also take great care when reproducing work for our baselines, making sure the experimental set ups are reasonable and match them as closely as we can to their descriptions.

\vspace{-0.1cm}
\section{Experiment Results}
This section provides the experiment results for the research questions we ask in Section \ref{sec:rqs}. To answer RQ$_1$ we use a combination of automated metrics and discuss its performance compared to other models in the context of these metrics. For RQ$_2$ we test a series of models with different hop sizes and compare them to our model as well as our baselines. To answer RQ$_3$ we provide a set of examples comparing our model using the graph AST and ConvGNN with a flattened AST model and show how the addition of the ConvGNN contributes to the overall summary.

\vspace{-0.1cm}
\subsection{RQ$_1$: Quantitative Evaluation}
\label{sec:rq1}

For our experimental results we tested three model configurations. In labeling our models we use code+gnn to represent the models that use an encoder for the source code tokens, a ConvGNN encoder for the AST, and then we use a +\{layer\_name\} format to show the layer that was used on the output of the ConvGNN.
% \vspace{-0.1cm}
% \begin{table}[!h]
% \vspace{-0.5cm}
% \begin{tabular}{c|c|c}
%   \textbf{code+gnn+dense}  & \textbf{code+gnn+BiLSTM}& \textbf{code+gnn+GRU}\\
% \end{tabular}
% \vspace{-0.5cm}
% \end{table}

We found that model code+gnn+BiLSTM was the highest performing approach obtaining a BLEU-A score of 19.93 and ROUGE-LCS score of 56.08, as seen in Table~\ref{tab:metrics}. The code+gnn+BiLSTM model outperformed the nearest graph-based baseline by 4.6\% BLEU-A and 0.06\% ROUGE-LCS. The code+gnn+BiLSTM model also outperformed the flattened AST baseline by 5.7\% BLEU-A and 12.72\% ROUGE-LCS. We attribute this increase in performance to the use of the ConvGNN as an encoding for the AST. Adding the ConvGNN allows the model to learn better AST node representations than it can with only a sequence model. We go into more depth into how the ConvGNN may be boosting performance in Section~\ref{sec:astexplain}. We attribute our performance improvement over other graph-based approaches to the use of the source code token sequence as a separate additional encoder. We found that using both the source code sequence and the AST allows the model to learn when to copy tokens directly from the source code, serving a purpose similar to a `copy mechanism' as described by Gu~\emph{et al.} \cite{gu2016copy}. Copy mechanisms are used to copy words directly from the input text to the output, primarily used to improve performance with rare or unknown tokens. In this case, the model has learned to copy tokens directly from the source code. This works well for source code summarization because of the large overlap in source code and summary vocabulary (over 94\%). In Section~\ref{sec:astexplain} example 1 we show how models that use both source code and AST input utilize the source code attention like a copy mechanism.

% \begin{table}[!b]
% \vspace{-0.4cm}
%     \centering
%     \begin{tabular}{|c|c||c|c|c|c||}
%     \hline
%     &reference&\multicolumn{2}{c|}{ast-attendgru}&\multicolumn{2}{c|}{code+gnn+gru}\\
%     \hline
%     length&count&BLEU-A&count&BLEU-A&count\\
%     \hline
%          3&6985&14.09&10005&14.58&9205\\
%          4&9238&19.88&14847&20.02&13579\\
%          5&8006&18.56&8020&19.33&8417\\
%          6&8943&21.36&12750&21.83&13166\\
%          7&10591&18.07&15018&18.68&15858\\
%          8&10497&21.00&10624&22.07&10973\\
%          9&10768&30.88&8402&31.39&8451\\
%          10&9713&17.62&5360&18.04&5088\\
%          11&7618&11.74&2997&12.32&3024\\
%          12&5862&9.48&1542&9.82&1383\\
%          \hline
%     \end{tabular}
%     \caption{Length of summary and associated counts reference and generated summaries. Model performance peaks at a length of  nine tokens in the output summary.}
%     \label{tab:comparemt}
%     \vspace{-0.4cm}
% \end{table}

We also see a noticeable effect on model performance based on the recurrent layer after the ConvGNNs. LeClair \emph{et al.} achieved 18.69 BLEU-A using only a recurrent layer to encode the flattened AST sequence, and without a recurrent layer the code+gnn+dense model achieves a 19.46 BLEU-A. In an effort to see how different recurrent layers affect the models performance, we trained a two hop model using GRU and another model using a BiLSTM as shown in Table~\ref{tab:metrics}. We found that code+gnn+BiLSTM outperformed code+gnn+GRU by 0.05 BLEU-A and 0.3 ROUGE-LCS. The improved score of the BiLSTM layer is likely due to the increased complexity of the layer over the GRU. We find in many cases that the BiLSTM architecture outperforms other recurrent layers, but at a significantly increased computational cost. For this reason, and because the code+gnn+BiLSTM model only outperformed the code+gnn+GRU model by 0.05 BLEU-A (0.2\%), we chose to conduct our other tests using the code+gnn+GRU architecture.

% Another observation we made is that, when comparing code+gnn+GRU to ast-attendgru, model code+gnn+GRU had higher BLEU-A scores for every length category. In other words, code+gnn+GRU is not outperforming ast-attendgru merely ebcause it does well on e.g. short or long output sequences. Consider Table~\ref{tab:comparemt}. Notice how the code+gnn+GRU model generated more sequences in the seven to nine token length range, which are the most common reference lengths, while the ast-attengru model generated more summaries of length 3 to 4, and of lengths 12. This shows that the code+gnn+GRU model learned to generate sequences better matching the reference texts, and of those generated, produced high quality summaries when scored by BLEU.

\vspace{-0.1cm}
\subsection{RQ$_2$: Hop size analysis}

In Table~\ref{tab:metrics} we compare the number of hops in the ConvGNN layers and how it affects the performance of the model. We found that for the AST two hops had the best overall performance. With having two hops, a node will get information from nodes up to two edges away. As outlined in Section~\ref{sec:approach}, our model implementation creates a separate ConvGNN layer for each hop in series. One explanation for why two hops had the best performance is that, because we are generating summaries at the method level, the ASTs in the dataset are not very deep. Another possibility could be that the other nodes most important to a specific node are its neighbors, and dealing with smaller clusters of node data is sufficient for learning. Lastly, even though the number of hops directly influences how far and quickly information will propagate through the ConvGNN, every iteration the neighboring nodes are now an aggregate of their neighbors $n$ hops away. In other words, after one iteration with two hops, a nodes neighbor is now an aggregate of nodes three hops away, so after enough iterations each node should be affected by every other node, with closer nodes having a larger effect.

While using two hops reported the best BLEU score for the code+gnn+GRU models, it only performed 1.5\% better than using three hops and 2.8\% better than using 10. Also notice that using five hops outperformed three and ten hops, this could be due to the random initialization or other minor factors. Because the difference in overall BLEU score is relatively small between hop sizes, we believe that the number of hops is less important than other hyper-parameters. It could be that the number of hops will be more important when summarizing larger selections of code where nodes are farther apart. For example, if the task were to summarize an entire program it may be beneficial to have more hops in your encoder to allow information to propagate farther.

% \begin{table}[!h]
% \subfloat[Example 1\label{tab:example2}]{

% }
% \end{table}
%\vspace{-0.3cm}

\vspace{-0.1cm}
\subsection{RQ$_3$: Graph AST Contribution}
\label{sec:astexplain}

\begin{figure}[t!]
\textbf{Example 1, Method ID 20477616}
\subfloat{\parbox{0.96\columnwidth}{
\vspace{-0.3cm}
\begin{tabular}{ll} %\hline
\hline
\textbf{summaries}&\\
\emph{reference}      & sends a guess to the server \\
\emph{code+gnn+GRU}  & sends a guess to the socket	\\
\emph{ast-attendgru} &attempts to initiate a <UNK> guess\\[5pt]
\hline
\vspace{0.1cm}
\textbf{source code}&\\
%\hline
\end{tabular}
}}
\vspace{0.05cm}
{\small
\begin{verbatim}
    public void sendGuess(String guess) {
      if( isConnected() ) {
        gui.statusBarInfo("Querying...", false);
        try {
          os.write( (guess + "\\r\\n").getBytes() );
          os.flush();
        } catch (IOException e) {
    gui.statusBarInfo(
        "Failed to send guess.IOException",true
        );
    System.err.println(
        "IOException during send guess to server"
        );
        }
      }
    }
\end{verbatim}
}
\vspace{0.1cm}

\centering
\subfloat[\emph{code+gnn+GRU}: Source attention]{
\parbox{\columnwidth}{
    \includegraphics[width=\columnwidth]{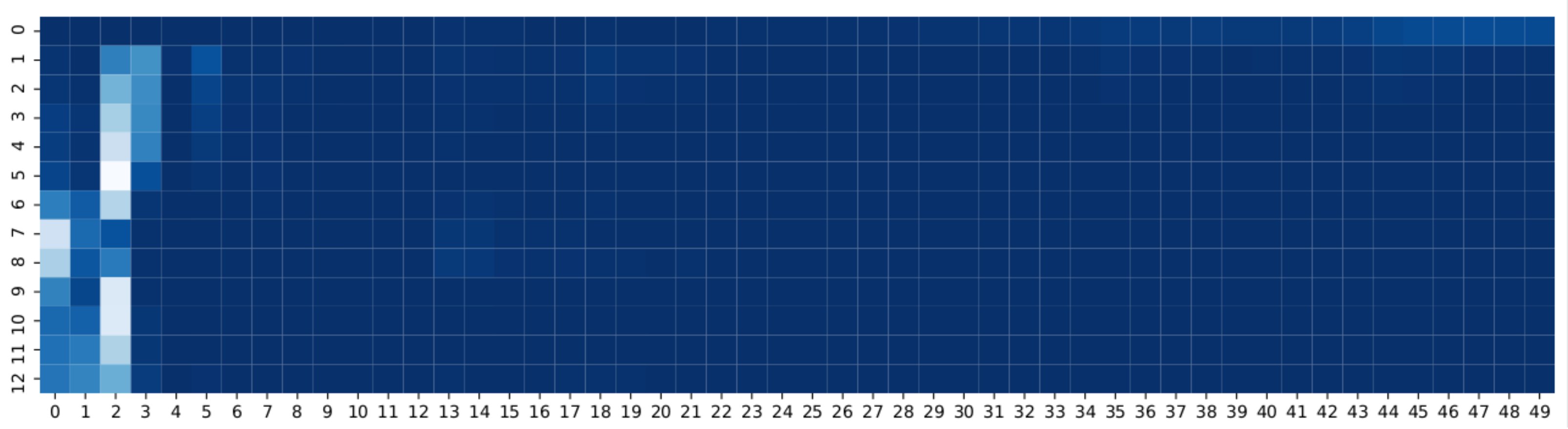}
}
}
\vspace{-0.3cm}
\subfloat[\emph{code+gnn+GRU}: AST attention]{
\parbox{\columnwidth}{
    \includegraphics[width=\columnwidth]{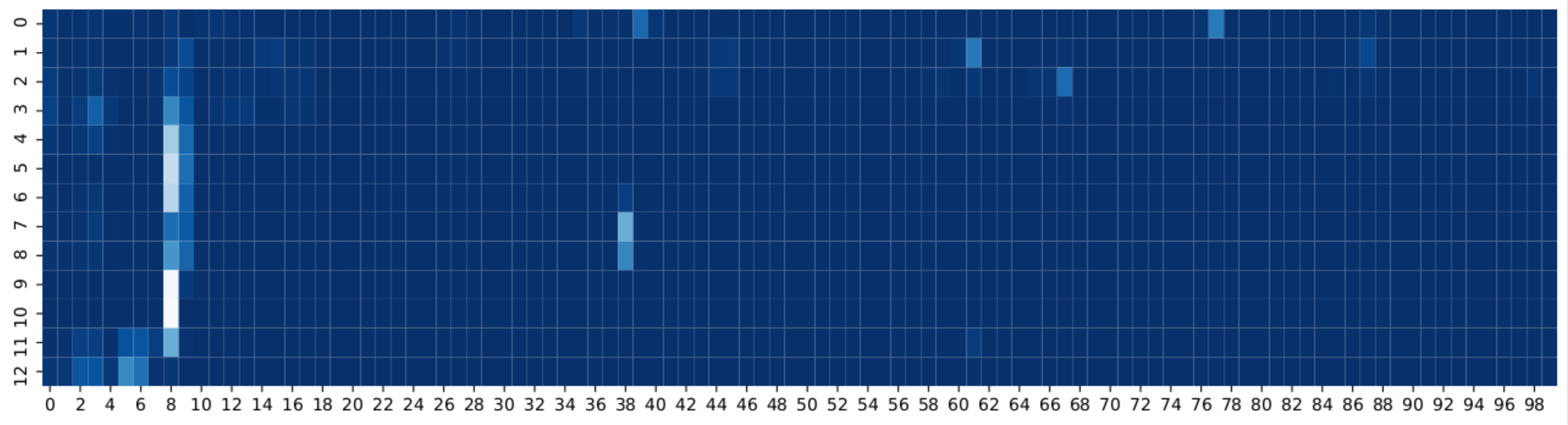}
}
}
\vspace{-0.3cm}
\subfloat[\emph{ast-attendgru}: Source attention]{
\parbox{\columnwidth}{
    \includegraphics[width=\columnwidth]{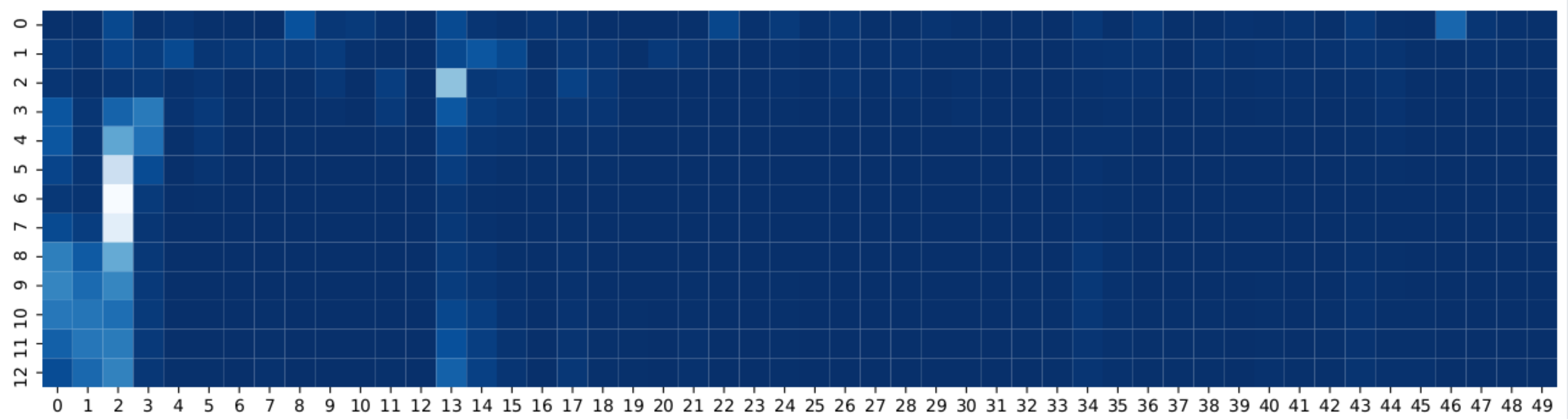}
}
}
\vspace{-0.3cm}
\subfloat[\emph{ast-attendgru}: AST attention]{
\parbox{\columnwidth}{
    \includegraphics[width=\columnwidth]{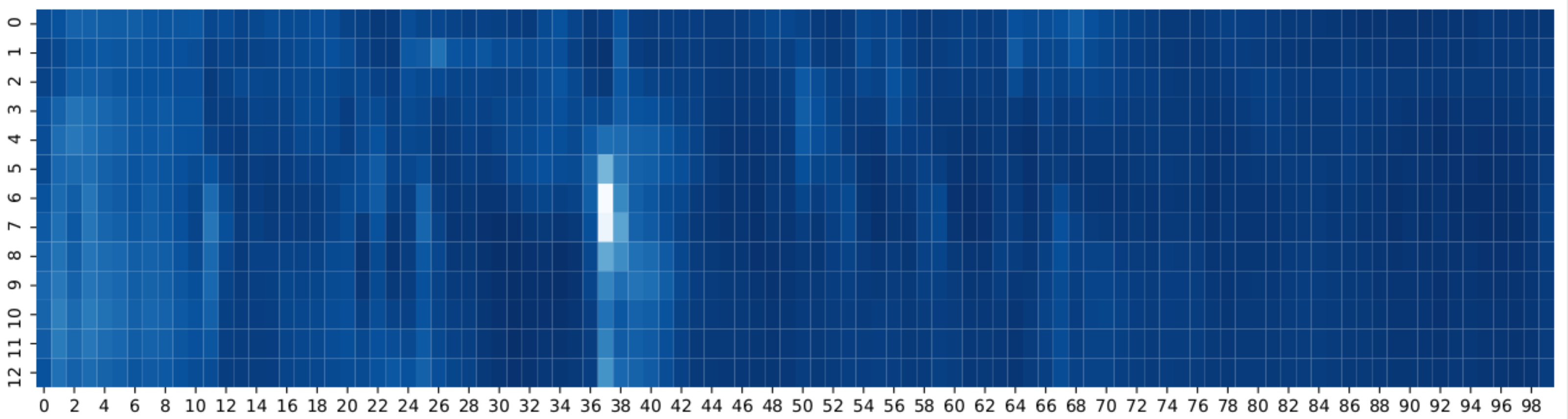}
}
}
\caption*{Example 1: Visualization of source code and AST attention for code+gnn+gru and ast-attendgru}
\label{fig:example1}
\end{figure}

We provide three in-depth examples of the GNN's contribution to our model. In these examples we also compare directly with the ast-attendgru model proposed by LeClair~\emph{et al.} \cite{leclair2019neural}. We chose to compare with ast-attendgru because it represents a collection of work using flattened ASTs to summarize source code as well as having separate encoders for the source code sequence and AST. We feel that comparing against this model allows us to isolate the contribution that the ConvGNN is making to the generated summaries. It should be noted however that these two models process the AST differently, they both use SrcML~\cite{collard2011lightweight} to generate ASTs, but ast-attendgru takes another step and additionally processes the AST using the SBT technique developed by Hu~\emph{et al.}. More details about how LeClair~\emph{et al.} process the AST can be found in Section~\ref{sec:baselines}.

Our first example visualized in Example 1 shows an instance where the reference summary contains tokens that also appear in the source code. We use this example to showcase how the source code and AST attentions work together as a copy mechanism to generate summaries. The visualizations in Example 1 are a snapshot of the attention outputs for both the source code and AST when the models are predicting the first token in the sequence, which is `sends' in the reference. We can see in Example 1 (a) that the code+gnn+GRU model and (c) the ast-attendgru model attend to the third token in the input source code sequence (column 2), which in this case is the token `send'. Where these two models differ, however, is what their respective ASTs are attending to (seen in (b) and (d)). In the case of code+gnn+GRU (b), the model is attending to the token `send' in column eight and the token `status' in column thirty-eight. On the other hand, The ast-attendgru (d) model is attending to column thirty-seven, which is the token `sexpr\_stmt'.

One explanation for this is that code+gnn+GRU is able to combine structural and code elements better than models that don't utilize a ConvGNN. In the context of this example what this means is that the AST token `send' in the code+gnn+GRU is a learned combination of the `send' node and its neighbors, which in the AST are nodes `name', `status', `bar', and `info'. The ast-attendgru model only sees the AST as a sequence of tokens, so when it attends to the token `sexpr\_stmt', its neighboring tokens are `sblock' and `sexpr'. Another observation from Example 1 (d) is that, generally, the ast-attendgru model activates more on the AST sequence than it does on the source code sequence, while code+gnn+GRU activates similarly on both attentions and focuses more on specific tokens. This could be due to the ConvGNN learning more specific structure information from the AST.

We also found that because the AST attention is more fine grained than the ast-attendgru attention, it learns whether to copy words directly from the source code better than the other model. In this case, because both the source code and AST attention focus on the same token, `send', it determined that `send' or a word very close to it (in this case `sends') should be the predicted token. If the source code and AST attention differ then the model will often times predict tokens that are not in the source code or AST.

If we look at later tokens in the predicted sequence, we see that code+gnn+GRU predicts the correct tokens until the last one. For the final token the reference token is `server' and code+gnn+GRU predicted `socket'. What is also interesting here is that ast-attendgru predicted the token `guess' which is in the reference summary and source code sequence. If we look at Example 2 we see the output of the AST attention mechanisms for both the code+gnn+GRU and ast-attendgru models during their prediction of the final token in the sequence. What this means is that code+gnn+GRU has the input sequence [sends, a, guess, to, the] and predicts `socket' and ast-attendgru has the sequence [attempts, to, initiate, a, <UNK>] and predicts `guess'. We do not include the source code visualization here because they were very similar to the visualizations in Example 1 (a) and (c). So, in the source code attention both models attend to the tokens `send' and `guess', but as we can see in the AST visualization, code+gnn+GRU is attending to the token in column forty-six - `querying'; and ast-attendgru is attending to a large, non-specific area in the structure of the AST. The code+gnn+GRU model has learned that the combination of the tokens `send', `guess', and the AST token `querying' lead to the prediction of the token `socket'. While this prediction was incorrect, the token `socket' is closely related to the term `server' in this context. Notice that the token `querying' is also in the source code, but neither model attends to it. As stated above, the source code attention is acting more as a copy mechanism and is attending to tokens that it believes should be the next predicted token, then it relies on the AST attention to add additional information for the final prediction.

\begin{figure}[b!]
\vspace{-0.6cm}
\centering
\subfloat[\emph{code+gnn+GRU}: AST attention]{
\parbox{\columnwidth}{
    \includegraphics[width=\columnwidth]{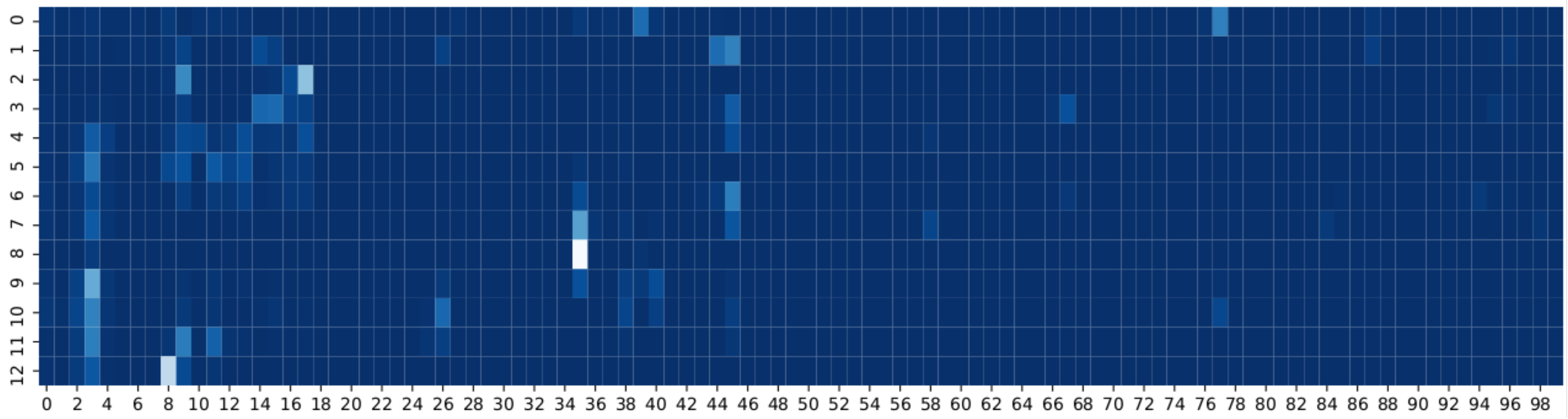}
}
}
\vspace{-0.3cm}
\subfloat[\emph{ast-attendgru}: AST attention]{
\parbox{\columnwidth}{
    \includegraphics[width=\columnwidth]{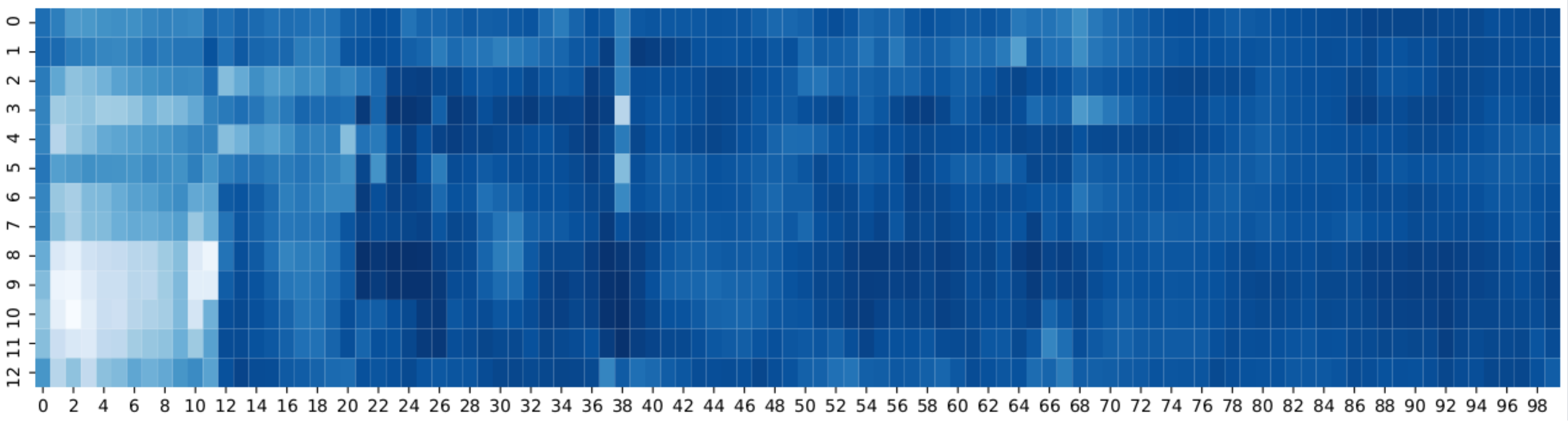}
}
}
\vspace{-0.1cm}
\caption*{Example 2: Visualization of AST attention mechanisms for code+gnn+gru and ast-attendgru when predicting the final token in the sequence.}
\label{fig:example1P2}
%\vspace{-0.6cm}
\end{figure}

Example 3 is a case where the code+gnn+GRU model correctly predicts the sixth token in the sequence, `first', but ast-attendgru predicts the token `specified'. For this prediction both models have the same predicted token sequence input to the decoder. If we look at Example 3, we can see the the attention visualizations for our models for their prediction of the sixth token in the sequence. In Example 3 (a) in the sixth row, the code+gnn+GRU model is attending to the fifth column, which is the token 'o'. In this piece of code 'o' is the identifier for the input parameter for the method. The ast-attendgru model source attention (Example 3 (c)) is attending to columns seventeen and thirty-four which are both the token `game'. Looking at the code+gnn+GRU AST attention (Example 3 (b)), we see that it is attending to column sixty-three, which is also the token `game'. So, in this example the ast-attengru model's source attention is attending to `game' and the code+gnn+GRU model's attention is also attending to the token `game' in the AST. This is important because it shows both models have learned that this token is important to the prediction, but in different contexts. Looking at the visualizations, we see again that the code+gnn+GRU model is able to focus on specific, important tokens in both the source code and the AST, while the ast-attendgru model attends to larger portions of the AST.

\vspace{-0.2cm}
\section{Discussion and Future Work}
The major take-aways from the work outlined in this paper are:
\vspace{-0.1cm}
\begin{itemize}
    \item Using the AST as a graph with ConvGNN layers outperforms a flattened version of the AST
    \item Including the source code sequence as a separate encoder allows the model to learn to use the source code and AST as a copy mechanism.
    \item The improved node embeddings from the ConvGNN allow the model to learn better token representations where representation of tokens in the AST are a combination of structure elements.
\end{itemize}
\vspace{-0.1cm}
The three examples that we show are situations where the addition of the ConvGNN allowed the model to learn better node representations than using a flattened sequence for the AST. When both the source code attention and the AST attention align on a specific token, the model treats this like a copy mechanism, directly copying the input source token to the output. When the source and AST attention do not agree, we see the model relying more on the AST to predict the next token in the sequence. When we compare this to a model with a flattened AST input, we see a large difference in how the AST is being attended to, generally the flattened AST model looks at larger structure areas instead of specific tokens. 

As an avenue for future work, models such as these have been shown to improve performance when ensembled. LeClair~\emph{et al.} showed that a model without AST information outperformed a model using AST information on specific types of summaries \cite{leclair2019neural}. This could lead to interesting results, potentially showing that bringing in different features from the source code allows the models to learn to generate better summaries for specific types of methods.

\vspace{-0.2cm}
\section{Conclusion}

In this work we have presented a new neural model architecture that utilizes a sequence of source code tokens along with ConvGNNs to encode the AST of a Java method and generate natural language summaries. We provide background and insights into why using a graph based neural network to encode the AST improves performance, along with providing a comparison of our results against relevant baselines. We conclude that the combination of source code tokens along with the AST and ConvGNNs allows the model to better learn when to directly copy tokens from the source code, as well as create better representations of tokens in the AST. We show that that the use of the ConvGNN to encode the AST improves aggregate BLEU scores (BLEU-A) by over 4.6\% over other graph-based approaches and 5.7\% improvement over flattened AST approaches. We also provide an in dept analysis of how the ConvGNN layers attribute to this increase in performance, and speculate on how these insights can be used for future work.

\begin{figure}[t!]
\textbf{Example 3, Method ID 25584536}
\subfloat{\parbox{0.96\columnwidth}{
\vspace{-0.4cm}
\begin{tabular}{ll} %\hline
\hline
\textbf{summaries}&\\
\emph{reference}      & returns the index of the first occurrence of\\ & the specified element \\
\emph{code+gnn+GRU}  & returns the index of the first occurrence of\\ & the specified element	\\
\emph{ast-attendgru} &returns the index of the specified object\\ & in the list\\[1pt]
\hline
\\
\textbf{source code}&\\
%\hline
\end{tabular}
}}
%\vspace{0.05cm}
{\small
\begin{verbatim} 
    public int indexOf(Object o) {
      if (o == null) {
        for (int i = 0; i < size; i++) {
          if (gameObjects[i] == null) {
            return i;
          }
        }
        } else {
            for (int i = 0; i < size; i++) {
                if (o.equals(gameObjects[i])) {
                    return i;
                }
            }
        }
        return -1;
    }
\end{verbatim}
}
    
\vspace{-0.3cm}
\subfloat[\emph{code+gnn+GRU}: Source attention]{
\parbox{\columnwidth}{
    \includegraphics[width=\columnwidth,height=2.1cm]{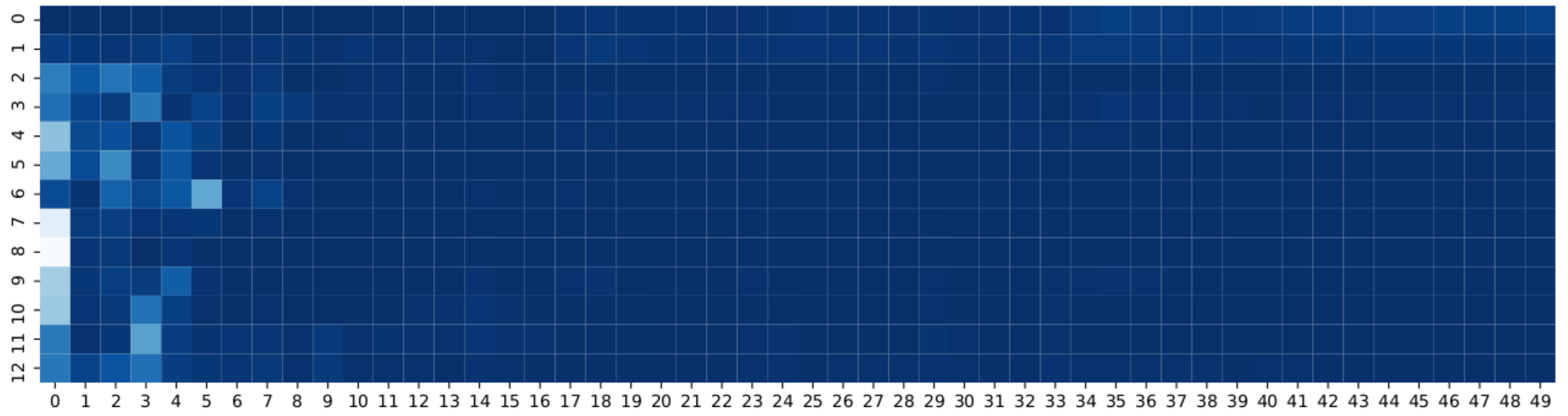}
}
}
\vspace{-0.3cm}
\subfloat[\emph{code+gnn+GRU}: AST attention]{
\parbox{\columnwidth}{
    \includegraphics[width=\columnwidth,height=2.1cm]{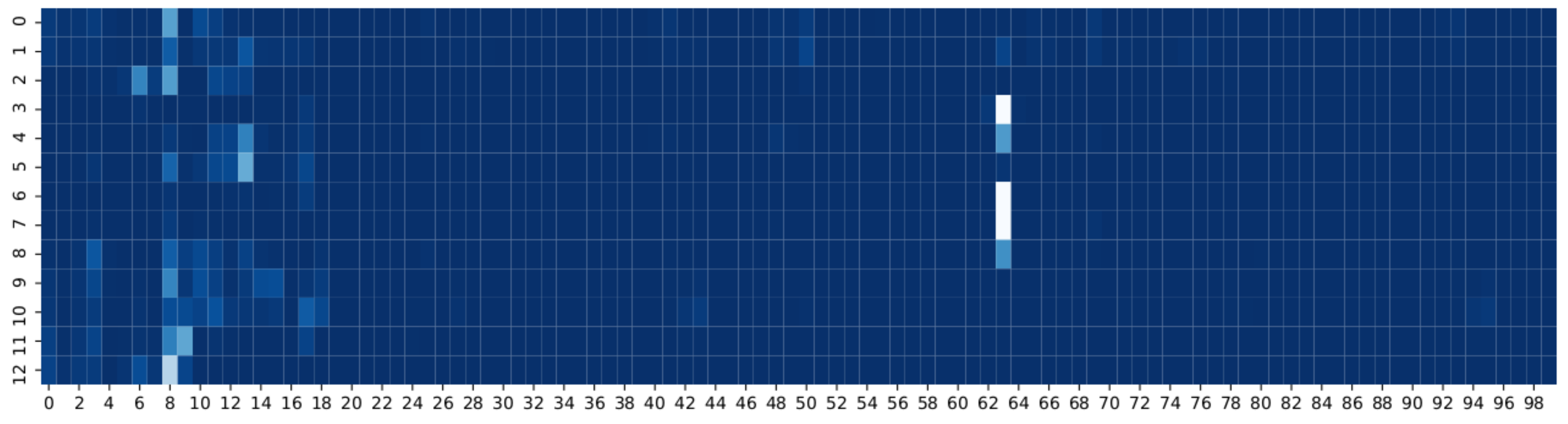}
}
}
\vspace{-0.3cm}
\subfloat[\emph{ast-attendgru}: Source attention]{
\parbox{\columnwidth}{
    \includegraphics[width=\columnwidth,height=2.1cm]{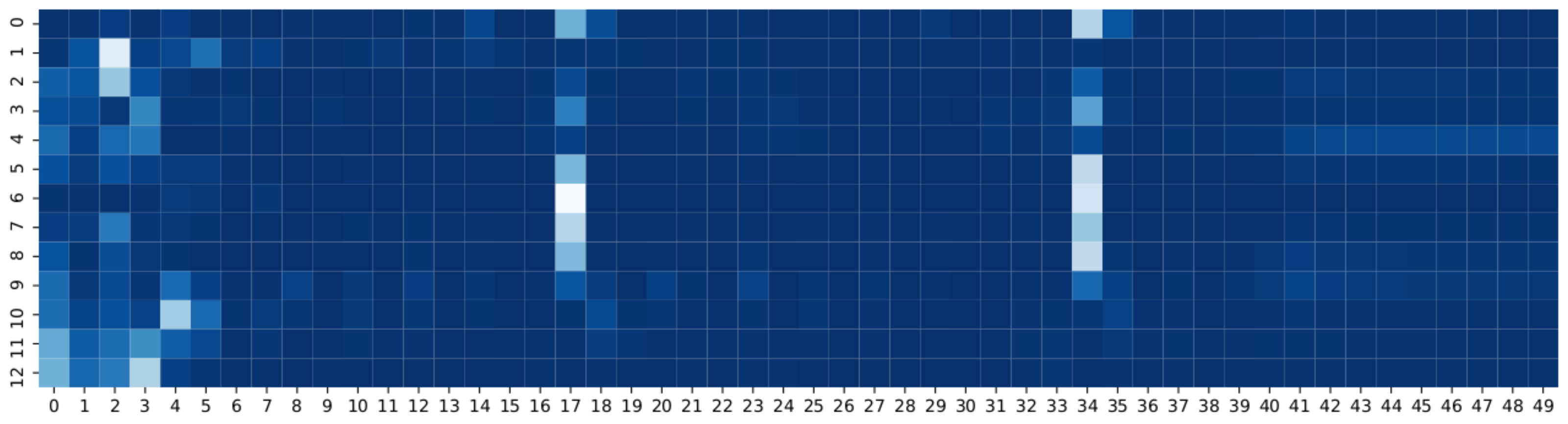}
}
}
\vspace{-0.3cm}
\subfloat[\emph{ast-attendgru}: AST attention]{
\parbox{\columnwidth}{
    \includegraphics[width=\columnwidth,height=2.1cm]{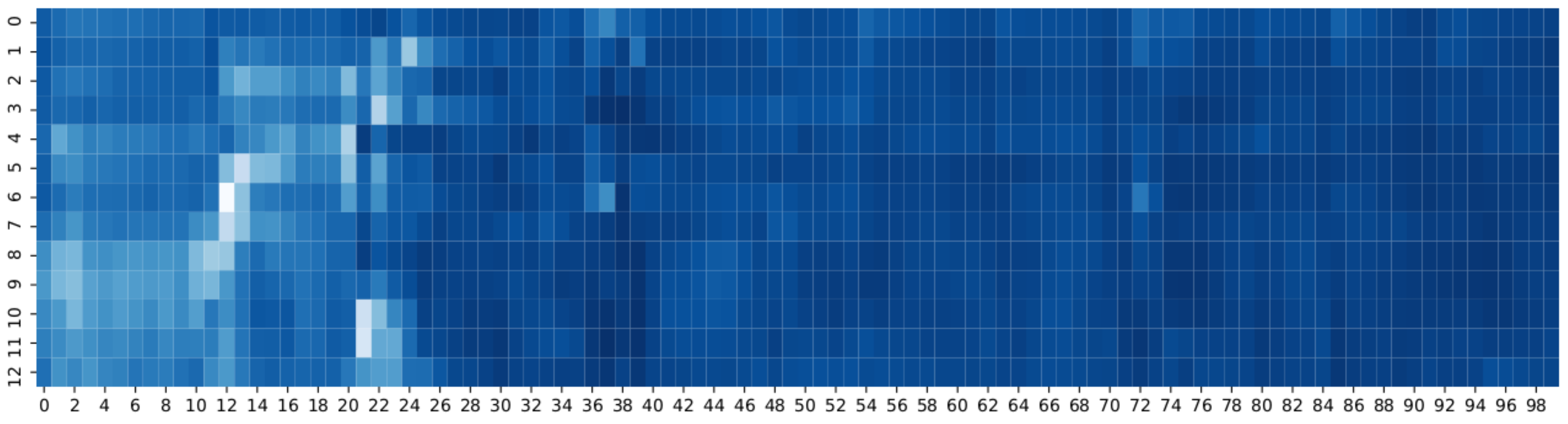}
}
}
\caption*{Example 3: Visualization of source code and AST attention for code+gnn+GRU and ast-attendgru}
\label{fig:example3}
%\vspace{0.2cm}
\end{figure}

\vspace{-0.2cm}
\section{Repoducibility}
\label{sec:repro}
All of our models, source code, and data used in this work can be found in our online repository at \url{https://go.aws/2tPXV2R}.

\vspace{-0.1cm}
\section{Acknowledgments}

\emph{This work is supported in part by NSF CCF-1452959 and CCF-1717607. Any opinions, findings, and conclusions expressed herein are the authors and do not necessarily reflect those of the sponsors.}

\bibliographystyle{ACM-Reference-Format}
\bibliography{main}

%%% -*-BibTeX-*-
%%% Do NOT edit. File created by BibTeX with style
%%% ACM-Reference-Format-Journals [18-Jan-2012].

\begin{thebibliography}{59}

%%% ====================================================================
%%% NOTE TO THE USER: you can override these defaults by providing
%%% customized versions of any of these macros before the \bibliography
%%% command.  Each of them MUST provide its own final punctuation,
%%% except for \shownote{}, \showDOI{}, and \showURL{}.  The latter two
%%% do not use final punctuation, in order to avoid confusing it with
%%% the Web address.
%%%
%%% To suppress output of a particular field, define its macro to expand
%%% to an empty string, or better, \unskip, like this:
%%%
%%% \newcommand{\showDOI}[1]{\unskip}   % LaTeX syntax
%%%
%%% \def \showDOI #1{\unskip}           % plain TeX syntax
%%%
%%% ====================================================================

\ifx \showCODEN    \undefined \def \showCODEN     #1{\unskip}     \fi
\ifx \showDOI      \undefined \def \showDOI       #1{#1}\fi
\ifx \showISBNx    \undefined \def \showISBNx     #1{\unskip}     \fi
\ifx \showISBNxiii \undefined \def \showISBNxiii  #1{\unskip}     \fi
\ifx \showISSN     \undefined \def \showISSN      #1{\unskip}     \fi
\ifx \showLCCN     \undefined \def \showLCCN      #1{\unskip}     \fi
\ifx \shownote     \undefined \def \shownote      #1{#1}          \fi
\ifx \showarticletitle \undefined \def \showarticletitle #1{#1}   \fi
\ifx \showURL      \undefined \def \showURL       {\relax}        \fi
% The following commands are used for tagged output and should be
% invisible to TeX
\providecommand\bibfield[2]{#2}
\providecommand\bibinfo[2]{#2}
\providecommand\natexlab[1]{#1}
\providecommand\showeprint[2][]{arXiv:#2}

\bibitem[\protect\citeauthoryear{Abadi, Agarwal, Barham, Brevdo, Chen, Citro,
  Corrado, Davis, Dean, Devin, Ghemawat, Goodfellow, Harp, Irving, Isard, Jia,
  Jozefowicz, Kaiser, Kudlur, Levenberg, Man\'{e}, Monga, Moore, Murray, Olah,
  Schuster, Shlens, Steiner, Sutskever, Talwar, Tucker, Vanhoucke, Vasudevan,
  Vi\'{e}gas, Vinyals, Warden, Wattenberg, Wicke, Yu, and Zheng}{Abadi
  et~al\mbox{.}}{2015}]%
        {tensorflow2015whitepaper}
\bibfield{author}{\bibinfo{person}{Mart\'{\i}n Abadi}, \bibinfo{person}{Ashish
  Agarwal}, \bibinfo{person}{Paul Barham}, \bibinfo{person}{Eugene Brevdo},
  \bibinfo{person}{Zhifeng Chen}, \bibinfo{person}{Craig Citro},
  \bibinfo{person}{Greg~S. Corrado}, \bibinfo{person}{Andy Davis},
  \bibinfo{person}{Jeffrey Dean}, \bibinfo{person}{Matthieu Devin},
  \bibinfo{person}{Sanjay Ghemawat}, \bibinfo{person}{Ian Goodfellow},
  \bibinfo{person}{Andrew Harp}, \bibinfo{person}{Geoffrey Irving},
  \bibinfo{person}{Michael Isard}, \bibinfo{person}{Yangqing Jia},
  \bibinfo{person}{Rafal Jozefowicz}, \bibinfo{person}{Lukasz Kaiser},
  \bibinfo{person}{Manjunath Kudlur}, \bibinfo{person}{Josh Levenberg},
  \bibinfo{person}{Dandelion Man\'{e}}, \bibinfo{person}{Rajat Monga},
  \bibinfo{person}{Sherry Moore}, \bibinfo{person}{Derek Murray},
  \bibinfo{person}{Chris Olah}, \bibinfo{person}{Mike Schuster},
  \bibinfo{person}{Jonathon Shlens}, \bibinfo{person}{Benoit Steiner},
  \bibinfo{person}{Ilya Sutskever}, \bibinfo{person}{Kunal Talwar},
  \bibinfo{person}{Paul Tucker}, \bibinfo{person}{Vincent Vanhoucke},
  \bibinfo{person}{Vijay Vasudevan}, \bibinfo{person}{Fernanda Vi\'{e}gas},
  \bibinfo{person}{Oriol Vinyals}, \bibinfo{person}{Pete Warden},
  \bibinfo{person}{Martin Wattenberg}, \bibinfo{person}{Martin Wicke},
  \bibinfo{person}{Yuan Yu}, {and} \bibinfo{person}{Xiaoqiang Zheng}.}
  \bibinfo{year}{2015}\natexlab{}.
\newblock \bibinfo{title}{{TensorFlow}: Large-Scale Machine Learning on
  Heterogeneous Systems}.
\newblock
\newblock
\urldef\tempurl%
\url{https://www.tensorflow.org/}
\showURL{%
\tempurl}
\newblock
\shownote{Software available from tensorflow.org.}


\bibitem[\protect\citeauthoryear{Allamanis, Brockschmidt, and
  Khademi}{Allamanis et~al\mbox{.}}{2018}]%
        {allamanis2018learning}
\bibfield{author}{\bibinfo{person}{Miltiadis Allamanis}, \bibinfo{person}{Marc
  Brockschmidt}, {and} \bibinfo{person}{Mahmoud Khademi}.}
  \bibinfo{year}{2018}\natexlab{}.
\newblock \showarticletitle{Learning to represent programs with graphs}.
\newblock \bibinfo{journal}{\emph{International Conference on Learning
  Representations}} (\bibinfo{year}{2018}).
\newblock


\bibitem[\protect\citeauthoryear{Alon, Brody, Levy, and Yahav}{Alon
  et~al\mbox{.}}{2019}]%
        {alon2018code2seq}
\bibfield{author}{\bibinfo{person}{Uri Alon}, \bibinfo{person}{Shaked Brody},
  \bibinfo{person}{Omer Levy}, {and} \bibinfo{person}{Eran Yahav}.}
  \bibinfo{year}{2019}\natexlab{}.
\newblock \showarticletitle{code2seq: Generating sequences from structured
  representations of code}.
\newblock \bibinfo{journal}{\emph{International Conference on Learning
  Representations}} (\bibinfo{year}{2019}).
\newblock


\bibitem[\protect\citeauthoryear{Arras, Horn, Montavon, Müller, and
  Samek}{Arras et~al\mbox{.}}{2017}]%
        {Arras2017explainabletext}
\bibfield{author}{\bibinfo{person}{Leila Arras}, \bibinfo{person}{Franziska
  Horn}, \bibinfo{person}{Grégoire Montavon}, \bibinfo{person}{Klaus-Robert
  Müller}, {and} \bibinfo{person}{Wojciech Samek}.}
  \bibinfo{year}{2017}\natexlab{}.
\newblock \showarticletitle{“What is relevant in a text document?”: An
  interpretable machine learning approach}.
\newblock \bibinfo{journal}{\emph{PLOS ONE}} \bibinfo{volume}{12},
  \bibinfo{number}{8} (\bibinfo{date}{Aug} \bibinfo{year}{2017}),
  \bibinfo{pages}{e0181142}.
\newblock
\showISSN{1932-6203}
\urldef\tempurl%
\url{https://doi.org/10.1371/journal.pone.0181142}
\showDOI{\tempurl}


\bibitem[\protect\citeauthoryear{Bahdanau, Cho, and Bengio}{Bahdanau
  et~al\mbox{.}}{2014}]%
        {bahdanau2014neural}
\bibfield{author}{\bibinfo{person}{Dzmitry Bahdanau},
  \bibinfo{person}{Kyunghyun Cho}, {and} \bibinfo{person}{Yoshua Bengio}.}
  \bibinfo{year}{2014}\natexlab{}.
\newblock \showarticletitle{Neural machine translation by jointly learning to
  align and translate}.
\newblock \bibinfo{journal}{\emph{arXiv preprint arXiv:1409.0473}}
  (\bibinfo{year}{2014}).
\newblock


\bibitem[\protect\citeauthoryear{Binkley}{Binkley}{2007}]%
        {binkley2007source}
\bibfield{author}{\bibinfo{person}{David Binkley}.}
  \bibinfo{year}{2007}\natexlab{}.
\newblock \showarticletitle{Source code analysis: A road map}. In
  \bibinfo{booktitle}{\emph{2007 Future of Software Engineering}}. IEEE
  Computer Society, \bibinfo{pages}{104--119}.
\newblock


\bibitem[\protect\citeauthoryear{Chen, Huang, Chiang, and Chen}{Chen
  et~al\mbox{.}}{2017}]%
        {chen2017improved}
\bibfield{author}{\bibinfo{person}{Huadong Chen}, \bibinfo{person}{Shujian
  Huang}, \bibinfo{person}{David Chiang}, {and} \bibinfo{person}{Jiajun Chen}.}
  \bibinfo{year}{2017}\natexlab{}.
\newblock \showarticletitle{Improved Neural Machine Translation with a
  Syntax-Aware Encoder and Decoder}. In \bibinfo{booktitle}{\emph{Proceedings
  of the 55th Annual Meeting of the Association for Computational Linguistics
  (Volume 1: Long Papers)}}. \bibinfo{publisher}{Association for Computational
  Linguistics}, \bibinfo{address}{Vancouver, Canada},
  \bibinfo{pages}{1936--1945}.
\newblock
\urldef\tempurl%
\url{https://doi.org/10.18653/v1/P17-1177}
\showDOI{\tempurl}


\bibitem[\protect\citeauthoryear{Chen, Wu, and Zaki}{Chen
  et~al\mbox{.}}{2020}]%
        {chen2019reinforcement}
\bibfield{author}{\bibinfo{person}{Yu Chen}, \bibinfo{person}{Lingfei Wu},
  {and} \bibinfo{person}{Mohammed~J Zaki}.} \bibinfo{year}{2020}\natexlab{}.
\newblock \showarticletitle{Reinforcement learning based graph-to-sequence
  model for natural question generation}.
\newblock \bibinfo{journal}{\emph{International Conference on Learning
  Representations}} (\bibinfo{year}{2020}).
\newblock


\bibitem[\protect\citeauthoryear{Chollet et~al\mbox{.}}{Chollet
  et~al\mbox{.}}{2015}]%
        {chollet2015keras}
\bibfield{author}{\bibinfo{person}{Fran\c{c}ois Chollet} {et~al\mbox{.}}}
  \bibinfo{year}{2015}\natexlab{}.
\newblock \bibinfo{title}{Keras}.
\newblock \bibinfo{howpublished}{\url{https://github.com/fchollet/keras}}.
\newblock


\bibitem[\protect\citeauthoryear{Collard, Decker, and Maletic}{Collard
  et~al\mbox{.}}{2011}]%
        {collard2011lightweight}
\bibfield{author}{\bibinfo{person}{Michael~L Collard},
  \bibinfo{person}{Michael~J Decker}, {and} \bibinfo{person}{Jonathan~I
  Maletic}.} \bibinfo{year}{2011}\natexlab{}.
\newblock \showarticletitle{Lightweight transformation and fact extraction with
  the srcML toolkit}. In \bibinfo{booktitle}{\emph{Source Code Analysis and
  Manipulation (SCAM), 2011 11th IEEE International Working Conference on}}.
  IEEE, \bibinfo{pages}{173--184}.
\newblock


\bibitem[\protect\citeauthoryear{Cornelissen, Zaidman, Van~Deursen, Moonen, and
  Koschke}{Cornelissen et~al\mbox{.}}{2009}]%
        {cornelissen2009systematic}
\bibfield{author}{\bibinfo{person}{Bas Cornelissen}, \bibinfo{person}{Andy
  Zaidman}, \bibinfo{person}{Arie Van~Deursen}, \bibinfo{person}{Leon Moonen},
  {and} \bibinfo{person}{Rainer Koschke}.} \bibinfo{year}{2009}\natexlab{}.
\newblock \showarticletitle{A systematic survey of program comprehension
  through dynamic analysis}.
\newblock \bibinfo{journal}{\emph{IEEE Transactions on Software Engineering}}
  \bibinfo{volume}{35}, \bibinfo{number}{5} (\bibinfo{year}{2009}),
  \bibinfo{pages}{684--702}.
\newblock


\bibitem[\protect\citeauthoryear{de~Souza, Anquetil, and de~Oliveira}{de~Souza
  et~al\mbox{.}}{2005}]%
        {deSouza:2005:SDE:1085313.1085331}
\bibfield{author}{\bibinfo{person}{Sergio Cozzetti~B. de Souza},
  \bibinfo{person}{Nicolas Anquetil}, {and} \bibinfo{person}{K\'{a}thia~M. de
  Oliveira}.} \bibinfo{year}{2005}\natexlab{}.
\newblock \showarticletitle{A study of the documentation essential to software
  maintenance}. In \bibinfo{booktitle}{\emph{Proceedings of the 23rd annual
  international conference on Design of communication: documenting \& designing
  for pervasive information}} \emph{(\bibinfo{series}{SIGDOC '05})}.
  \bibinfo{publisher}{ACM}, \bibinfo{address}{New York, NY, USA},
  \bibinfo{pages}{68--75}.
\newblock
\showISBNx{1-59593-175-9}
\urldef\tempurl%
\url{https://doi.org/10.1145/1085313.1085331}
\showDOI{\tempurl}


\bibitem[\protect\citeauthoryear{Doran, Schulz, and Besold}{Doran
  et~al\mbox{.}}{2017}]%
        {doran2017explainable}
\bibfield{author}{\bibinfo{person}{Derek Doran}, \bibinfo{person}{Sarah
  Schulz}, {and} \bibinfo{person}{Tarek~R. Besold}.}
  \bibinfo{year}{2017}\natexlab{}.
\newblock \showarticletitle{What Does Explainable {AI} Really Mean? {A} New
  Conceptualization of Perspectives}.
\newblock \bibinfo{journal}{\emph{CoRR}}  \bibinfo{volume}{abs/1710.00794}
  (\bibinfo{year}{2017}).
\newblock
\showeprint[arxiv]{1710.00794}
\urldef\tempurl%
\url{http://arxiv.org/abs/1710.00794}
\showURL{%
\tempurl}


\bibitem[\protect\citeauthoryear{{Doshi-Velez} and {Kim}}{{Doshi-Velez} and
  {Kim}}{2017}]%
        {doshi2017interpretableml}
\bibfield{author}{\bibinfo{person}{Finale {Doshi-Velez}} {and}
  \bibinfo{person}{Been {Kim}}.} \bibinfo{year}{2017}\natexlab{}.
\newblock \showarticletitle{{Towards A Rigorous Science of Interpretable
  Machine Learning}}.
\newblock \bibinfo{journal}{\emph{arXiv e-prints}}, Article
  \bibinfo{articleno}{arXiv:1702.08608} (\bibinfo{date}{Feb}
  \bibinfo{year}{2017}), \bibinfo{numpages}{arXiv:1702.08608}~pages.
\newblock
\showeprint[arxiv]{stat.ML/1702.08608}


\bibitem[\protect\citeauthoryear{Eddy, Robinson, Kraft, and Carver}{Eddy
  et~al\mbox{.}}{2013}]%
        {eddy2013evaluating}
\bibfield{author}{\bibinfo{person}{Brian~P Eddy}, \bibinfo{person}{Jeffrey~A
  Robinson}, \bibinfo{person}{Nicholas~A Kraft}, {and}
  \bibinfo{person}{Jeffrey~C Carver}.} \bibinfo{year}{2013}\natexlab{}.
\newblock \showarticletitle{Evaluating source code summarization techniques:
  Replication and expansion}. In \bibinfo{booktitle}{\emph{Program
  Comprehension (ICPC), 2013 IEEE 21st International Conference on}}. IEEE,
  \bibinfo{pages}{13--22}.
\newblock


\bibitem[\protect\citeauthoryear{Fernandes, Allamanis, and
  Brockschmidt}{Fernandes et~al\mbox{.}}{2018}]%
        {fernandes2019structured}
\bibfield{author}{\bibinfo{person}{Patrick Fernandes},
  \bibinfo{person}{Miltiadis Allamanis}, {and} \bibinfo{person}{Marc
  Brockschmidt}.} \bibinfo{year}{2018}\natexlab{}.
\newblock \showarticletitle{Structured Neural Summarization}.
\newblock \bibinfo{journal}{\emph{CoRR}}  \bibinfo{volume}{abs/1811.01824}
  (\bibinfo{year}{2018}).
\newblock
\showeprint[arxiv]{1811.01824}
\urldef\tempurl%
\url{http://arxiv.org/abs/1811.01824}
\showURL{%
\tempurl}


\bibitem[\protect\citeauthoryear{Forward and Lethbridge}{Forward and
  Lethbridge}{2002}]%
        {Forward:2002:RSD:585058.585065}
\bibfield{author}{\bibinfo{person}{Andrew Forward} {and}
  \bibinfo{person}{Timothy~C. Lethbridge}.} \bibinfo{year}{2002}\natexlab{}.
\newblock \showarticletitle{The relevance of software documentation, tools and
  technologies: a survey}. In \bibinfo{booktitle}{\emph{Proceedings of the 2002
  ACM symposium on Document engineering}} \emph{(\bibinfo{series}{DocEng
  '02})}. \bibinfo{publisher}{ACM}, \bibinfo{address}{New York, NY, USA},
  \bibinfo{pages}{26--33}.
\newblock
\showISBNx{1-58113-594-7}
\urldef\tempurl%
\url{https://doi.org/10.1145/585058.585065}
\showDOI{\tempurl}


\bibitem[\protect\citeauthoryear{Gu, Lu, Li, and Li}{Gu et~al\mbox{.}}{2016}]%
        {gu2016copy}
\bibfield{author}{\bibinfo{person}{Jiatao Gu}, \bibinfo{person}{Zhengdong Lu},
  \bibinfo{person}{Hang Li}, {and} \bibinfo{person}{Victor~O.K. Li}.}
  \bibinfo{year}{2016}\natexlab{}.
\newblock \showarticletitle{Incorporating Copying Mechanism in
  Sequence-to-Sequence Learning}.
\newblock \bibinfo{journal}{\emph{Proceedings of the 54th Annual Meeting of the
  Association for Computational Linguistics (Volume 1: Long Papers)}}
  (\bibinfo{year}{2016}).
\newblock
\urldef\tempurl%
\url{https://doi.org/10.18653/v1/p16-1154}
\showDOI{\tempurl}


\bibitem[\protect\citeauthoryear{Haiduc, Aponte, Moreno, and Marcus}{Haiduc
  et~al\mbox{.}}{2010}]%
        {haiduc2010use}
\bibfield{author}{\bibinfo{person}{Sonia Haiduc}, \bibinfo{person}{Jairo
  Aponte}, \bibinfo{person}{Laura Moreno}, {and} \bibinfo{person}{Andrian
  Marcus}.} \bibinfo{year}{2010}\natexlab{}.
\newblock \showarticletitle{On the use of automated text summarization
  techniques for summarizing source code}. In \bibinfo{booktitle}{\emph{Reverse
  Engineering (WCRE), 2010 17th Working Conference on}}. IEEE,
  \bibinfo{pages}{35--44}.
\newblock


\bibitem[\protect\citeauthoryear{Haiduc and Marcus}{Haiduc and Marcus}{2008}]%
        {Haiduc:S:ICPC:2008}
\bibfield{author}{\bibinfo{person}{S. Haiduc} {and} \bibinfo{person}{A.
  Marcus}.} \bibinfo{year}{2008}\natexlab{}.
\newblock \showarticletitle{On the Use of Domain Terms in Source Code}. In
  \bibinfo{booktitle}{\emph{16th IEEE International Conference on Program
  Comprehension (ICPC'08)}}. \bibinfo{address}{Amsterdam, The Netherlands},
  \bibinfo{pages}{113--122}.
\newblock


\bibitem[\protect\citeauthoryear{Hellendoorn and Devanbu}{Hellendoorn and
  Devanbu}{2017}]%
        {hellendoorn2017deep}
\bibfield{author}{\bibinfo{person}{Vincent~J Hellendoorn} {and}
  \bibinfo{person}{Premkumar Devanbu}.} \bibinfo{year}{2017}\natexlab{}.
\newblock \showarticletitle{Are deep neural networks the best choice for
  modeling source code?}. In \bibinfo{booktitle}{\emph{Proceedings of the 2017
  11th Joint Meeting on Foundations of Software Engineering}}. ACM,
  \bibinfo{pages}{763--773}.
\newblock


\bibitem[\protect\citeauthoryear{{Howard}, {Gupta}, {Pollock}, and
  {Vijay-Shanker}}{{Howard} et~al\mbox{.}}{2013}]%
        {howard2013mappings}
\bibfield{author}{\bibinfo{person}{M.~J. {Howard}}, \bibinfo{person}{S.
  {Gupta}}, \bibinfo{person}{L. {Pollock}}, {and} \bibinfo{person}{K.
  {Vijay-Shanker}}.} \bibinfo{year}{2013}\natexlab{}.
\newblock \showarticletitle{Automatically mining software-based,
  semantically-similar words from comment-code mappings}. In
  \bibinfo{booktitle}{\emph{2013 10th Working Conference on Mining Software
  Repositories (MSR)}}. \bibinfo{pages}{377--386}.
\newblock
\showISSN{2160-1852}
\urldef\tempurl%
\url{https://doi.org/10.1109/MSR.2013.6624052}
\showDOI{\tempurl}


\bibitem[\protect\citeauthoryear{Hu, Li, Xia, Lo, and Jin}{Hu
  et~al\mbox{.}}{2018a}]%
        {hu2018deep}
\bibfield{author}{\bibinfo{person}{Xing Hu}, \bibinfo{person}{Ge Li},
  \bibinfo{person}{Xin Xia}, \bibinfo{person}{David Lo}, {and}
  \bibinfo{person}{Zhi Jin}.} \bibinfo{year}{2018}\natexlab{a}.
\newblock \showarticletitle{Deep code comment generation}. In
  \bibinfo{booktitle}{\emph{Proceedings of the 26th International Conference on
  Program Comprehension}}. ACM, \bibinfo{pages}{200--210}.
\newblock


\bibitem[\protect\citeauthoryear{Hu, Li, Xia, Lo, Lu, and Jin}{Hu
  et~al\mbox{.}}{2018b}]%
        {hu2018summarizing}
\bibfield{author}{\bibinfo{person}{Xing Hu}, \bibinfo{person}{Ge Li},
  \bibinfo{person}{Xin Xia}, \bibinfo{person}{David Lo}, \bibinfo{person}{Shuai
  Lu}, {and} \bibinfo{person}{Zhi Jin}.} \bibinfo{year}{2018}\natexlab{b}.
\newblock \showarticletitle{Summarizing Source Code with Transferred API
  Knowledge.}. In \bibinfo{booktitle}{\emph{IJCAI}}.
  \bibinfo{pages}{2269--2275}.
\newblock


\bibitem[\protect\citeauthoryear{Iyer, Konstas, Cheung, and Zettlemoyer}{Iyer
  et~al\mbox{.}}{2016}]%
        {iyer2016summarizing}
\bibfield{author}{\bibinfo{person}{Srinivasan Iyer}, \bibinfo{person}{Ioannis
  Konstas}, \bibinfo{person}{Alvin Cheung}, {and} \bibinfo{person}{Luke
  Zettlemoyer}.} \bibinfo{year}{2016}\natexlab{}.
\newblock \showarticletitle{Summarizing source code using a neural attention
  model}. In \bibinfo{booktitle}{\emph{Proceedings of the 54th Annual Meeting
  of the Association for Computational Linguistics (Volume 1: Long Papers)}},
  Vol.~\bibinfo{volume}{1}. \bibinfo{pages}{2073--2083}.
\newblock


\bibitem[\protect\citeauthoryear{Kajko-Mattsson}{Kajko-Mattsson}{2005}]%
        {Kajko-Mattsson:2005:SDP:1032622.1035374}
\bibfield{author}{\bibinfo{person}{Mira Kajko-Mattsson}.}
  \bibinfo{year}{2005}\natexlab{}.
\newblock \showarticletitle{A Survey of Documentation Practice within
  Corrective Maintenance}.
\newblock \bibinfo{journal}{\emph{Empirical Softw. Engg.}}
  \bibinfo{volume}{10}, \bibinfo{number}{1} (\bibinfo{date}{Jan.}
  \bibinfo{year}{2005}), \bibinfo{pages}{31--55}.
\newblock
\showISSN{1382-3256}
\urldef\tempurl%
\url{https://doi.org/10.1023/B:LIDA.0000048322.42751.ca}
\showDOI{\tempurl}


\bibitem[\protect\citeauthoryear{Kramer}{Kramer}{1999}]%
        {kramer1999api}
\bibfield{author}{\bibinfo{person}{Douglas Kramer}.}
  \bibinfo{year}{1999}\natexlab{}.
\newblock \showarticletitle{API documentation from source code comments: a case
  study of Javadoc}. In \bibinfo{booktitle}{\emph{Proceedings of the 17th
  annual international conference on Computer documentation}}. ACM,
  \bibinfo{pages}{147--153}.
\newblock


\bibitem[\protect\citeauthoryear{{LeClair}, {Eberhart}, and
  {McMillan}}{{LeClair} et~al\mbox{.}}{2018}]%
        {leclair2018codecat}
\bibfield{author}{\bibinfo{person}{A. {LeClair}}, \bibinfo{person}{Z.
  {Eberhart}}, {and} \bibinfo{person}{C. {McMillan}}.}
  \bibinfo{year}{2018}\natexlab{}.
\newblock \showarticletitle{Adapting Neural Text Classification for Improved
  Software Categorization}. In \bibinfo{booktitle}{\emph{2018 IEEE
  International Conference on Software Maintenance and Evolution (ICSME)}}.
  \bibinfo{pages}{461--472}.
\newblock
\showISSN{2576-3148}
\urldef\tempurl%
\url{https://doi.org/10.1109/ICSME.2018.00056}
\showDOI{\tempurl}


\bibitem[\protect\citeauthoryear{LeClair, Jiang, and McMillan}{LeClair
  et~al\mbox{.}}{2019}]%
        {leclair2019neural}
\bibfield{author}{\bibinfo{person}{Alexander LeClair}, \bibinfo{person}{Siyuan
  Jiang}, {and} \bibinfo{person}{Collin McMillan}.}
  \bibinfo{year}{2019}\natexlab{}.
\newblock \showarticletitle{A neural model for generating natural language
  summaries of program subroutines}. In \bibinfo{booktitle}{\emph{Proceedings
  of the 41st International Conference on Software Engineering}}. IEEE Press,
  \bibinfo{pages}{795--806}.
\newblock


\bibitem[\protect\citeauthoryear{LeClair and McMillan}{LeClair and
  McMillan}{2019}]%
        {leclair2019recommendations}
\bibfield{author}{\bibinfo{person}{Alexander LeClair} {and}
  \bibinfo{person}{Collin McMillan}.} \bibinfo{year}{2019}\natexlab{}.
\newblock \showarticletitle{Recommendations for Datasets for Source Code
  Summarization}. In \bibinfo{booktitle}{\emph{Proceedings of the 2019
  Conference of the North American Chapter of the Association for Computational
  Linguistics: Human Language Technologies, Volume 1 (Long and Short Papers)}}.
  \bibinfo{pages}{3931--3937}.
\newblock


\bibitem[\protect\citeauthoryear{Liang and Zhu}{Liang and Zhu}{2018}]%
        {liang2018automatic}
\bibfield{author}{\bibinfo{person}{Yuding Liang} {and}
  \bibinfo{person}{Kenny~Q. Zhu}.} \bibinfo{year}{2018}\natexlab{}.
\newblock \showarticletitle{Automatic Generation of Text Descriptive Comments
  for Code Blocks}.
\newblock \bibinfo{journal}{\emph{CoRR}}  \bibinfo{volume}{abs/1808.06880}
  (\bibinfo{year}{2018}).
\newblock
\showeprint[arxiv]{1808.06880}
\urldef\tempurl%
\url{http://arxiv.org/abs/1808.06880}
\showURL{%
\tempurl}


\bibitem[\protect\citeauthoryear{Lin}{Lin}{2004}]%
        {lin2004rouge}
\bibfield{author}{\bibinfo{person}{Chin-Yew Lin}.}
  \bibinfo{year}{2004}\natexlab{}.
\newblock \showarticletitle{Rouge: A package for automatic evaluation of
  summaries}.
\newblock \bibinfo{journal}{\emph{Text Summarization Branches Out}}
  (\bibinfo{year}{2004}).
\newblock


\bibitem[\protect\citeauthoryear{Lopes, Bajracharya, Ossher, and Baldi}{Lopes
  et~al\mbox{.}}{2010}]%
        {Lopes+Bajracharya+Ossher+Baldi:2010}
\bibfield{author}{\bibinfo{person}{C. Lopes}, \bibinfo{person}{S. Bajracharya},
  \bibinfo{person}{J. Ossher}, {and} \bibinfo{person}{P. Baldi}.}
  \bibinfo{year}{2010}\natexlab{}.
\newblock \bibinfo{title}{{UCI} Source Code Data Sets}.
\newblock
\newblock
\urldef\tempurl%
\url{http://www.ics.uci.edu/$\sim$lopes/datasets/}
\showURL{%
\tempurl}


\bibitem[\protect\citeauthoryear{Loyola, Marrese-Taylor, and Matsuo}{Loyola
  et~al\mbox{.}}{2017}]%
        {Loyola2017ANA}
\bibfield{author}{\bibinfo{person}{Pablo Loyola}, \bibinfo{person}{Edison
  Marrese-Taylor}, {and} \bibinfo{person}{Yutaka Matsuo}.}
  \bibinfo{year}{2017}\natexlab{}.
\newblock \showarticletitle{A Neural Architecture for Generating Natural
  Language Descriptions from Source Code Changes}. In
  \bibinfo{booktitle}{\emph{ACL}}.
\newblock


\bibitem[\protect\citeauthoryear{Lu, Zhao, Li, and Jin}{Lu
  et~al\mbox{.}}{2019}]%
        {lu2017learning}
\bibfield{author}{\bibinfo{person}{Yangyang Lu}, \bibinfo{person}{Zelong Zhao},
  \bibinfo{person}{Ge Li}, {and} \bibinfo{person}{Zhi Jin}.}
  \bibinfo{year}{2019}\natexlab{}.
\newblock \showarticletitle{Learning to Generate Comments for API-Based Code
  Snippets}. In \bibinfo{booktitle}{\emph{Software Engineering and Methodology
  for Emerging Domains}}, \bibfield{editor}{\bibinfo{person}{Zheng Li},
  \bibinfo{person}{He~Jiang}, \bibinfo{person}{Ge~Li}, \bibinfo{person}{Minghui
  Zhou}, {and} \bibinfo{person}{Ming Li}} (Eds.). \bibinfo{publisher}{Springer
  Singapore}, \bibinfo{address}{Singapore}, \bibinfo{pages}{3--14}.
\newblock
\showISBNx{978-981-15-0310-8}


\bibitem[\protect\citeauthoryear{McBurney, Liu, and McMillan}{McBurney
  et~al\mbox{.}}{2016}]%
        {mcburney2016automated}
\bibfield{author}{\bibinfo{person}{Paul~W McBurney}, \bibinfo{person}{Cheng
  Liu}, {and} \bibinfo{person}{Collin McMillan}.}
  \bibinfo{year}{2016}\natexlab{}.
\newblock \showarticletitle{Automated feature discovery via sentence selection
  and source code summarization}.
\newblock \bibinfo{journal}{\emph{Journal of Software: Evolution and Process}}
  \bibinfo{volume}{28}, \bibinfo{number}{2} (\bibinfo{year}{2016}),
  \bibinfo{pages}{120--145}.
\newblock


\bibitem[\protect\citeauthoryear{McBurney and McMillan}{McBurney and
  McMillan}{2016}]%
        {mcburney2016automatic}
\bibfield{author}{\bibinfo{person}{Paul~W McBurney} {and}
  \bibinfo{person}{Collin McMillan}.} \bibinfo{year}{2016}\natexlab{}.
\newblock \showarticletitle{Automatic source code summarization of context for
  java methods}.
\newblock \bibinfo{journal}{\emph{IEEE Transactions on Software Engineering}}
  \bibinfo{volume}{42}, \bibinfo{number}{2} (\bibinfo{year}{2016}),
  \bibinfo{pages}{103--119}.
\newblock


\bibitem[\protect\citeauthoryear{Miller}{Miller}{2019}]%
        {MILLER2019explainai}
\bibfield{author}{\bibinfo{person}{Tim Miller}.}
  \bibinfo{year}{2019}\natexlab{}.
\newblock \showarticletitle{Explanation in artificial intelligence: Insights
  from the social sciences}.
\newblock \bibinfo{journal}{\emph{Artificial Intelligence}}
  \bibinfo{volume}{267} (\bibinfo{year}{2019}), \bibinfo{pages}{1 -- 38}.
\newblock
\showISSN{0004-3702}
\urldef\tempurl%
\url{https://doi.org/10.1016/j.artint.2018.07.007}
\showDOI{\tempurl}


\bibitem[\protect\citeauthoryear{Moreno and Aponte}{Moreno and Aponte}{2012}]%
        {moreno2012analysis}
\bibfield{author}{\bibinfo{person}{Laura Moreno} {and} \bibinfo{person}{Jairo
  Aponte}.} \bibinfo{year}{2012}\natexlab{}.
\newblock \showarticletitle{On the analysis of human and automatic summaries of
  source code}.
\newblock \bibinfo{journal}{\emph{CLEI Electronic Journal}}
  \bibinfo{volume}{15}, \bibinfo{number}{2} (\bibinfo{year}{2012}),
  \bibinfo{pages}{2--2}.
\newblock


\bibitem[\protect\citeauthoryear{Moreno, Aponte, Sridhara, Marcus, Pollock, and
  Vijay-Shanker}{Moreno et~al\mbox{.}}{2013}]%
        {moreno2013automatic}
\bibfield{author}{\bibinfo{person}{Laura Moreno}, \bibinfo{person}{Jairo
  Aponte}, \bibinfo{person}{Giriprasad Sridhara}, \bibinfo{person}{Andrian
  Marcus}, \bibinfo{person}{Lori Pollock}, {and} \bibinfo{person}{K
  Vijay-Shanker}.} \bibinfo{year}{2013}\natexlab{}.
\newblock \showarticletitle{Automatic generation of natural language summaries
  for java classes}. In \bibinfo{booktitle}{\emph{Program Comprehension (ICPC),
  2013 IEEE 21st International Conference on}}. IEEE, \bibinfo{pages}{23--32}.
\newblock


\bibitem[\protect\citeauthoryear{Nazar, Hu, and Jiang}{Nazar
  et~al\mbox{.}}{2016}]%
        {nazar2016summarizing}
\bibfield{author}{\bibinfo{person}{Najam Nazar}, \bibinfo{person}{Yan Hu},
  {and} \bibinfo{person}{He Jiang}.} \bibinfo{year}{2016}\natexlab{}.
\newblock \showarticletitle{Summarizing software artifacts: A literature
  review}.
\newblock \bibinfo{journal}{\emph{Journal of Computer Science and Technology}}
  \bibinfo{volume}{31}, \bibinfo{number}{5} (\bibinfo{year}{2016}),
  \bibinfo{pages}{883--909}.
\newblock


\bibitem[\protect\citeauthoryear{Ottenstein and Ottenstein}{Ottenstein and
  Ottenstein}{1984}]%
        {ottenstein1984program}
\bibfield{author}{\bibinfo{person}{Karl~J Ottenstein} {and}
  \bibinfo{person}{Linda~M Ottenstein}.} \bibinfo{year}{1984}\natexlab{}.
\newblock \showarticletitle{The program dependence graph in a software
  development environment}.
\newblock \bibinfo{journal}{\emph{ACM SIGSOFT Software Engineering Notes}}
  \bibinfo{volume}{9}, \bibinfo{number}{3} (\bibinfo{year}{1984}),
  \bibinfo{pages}{177--184}.
\newblock


\bibitem[\protect\citeauthoryear{Papineni, Roukos, Ward, and Zhu}{Papineni
  et~al\mbox{.}}{2002}]%
        {papineni2002bleu}
\bibfield{author}{\bibinfo{person}{Kishore Papineni}, \bibinfo{person}{Salim
  Roukos}, \bibinfo{person}{Todd Ward}, {and} \bibinfo{person}{Wei-Jing Zhu}.}
  \bibinfo{year}{2002}\natexlab{}.
\newblock \showarticletitle{BLEU: a method for automatic evaluation of machine
  translation}. In \bibinfo{booktitle}{\emph{Proceedings of the 40th annual
  meeting on association for computational linguistics}}. Association for
  Computational Linguistics, \bibinfo{pages}{311--318}.
\newblock


\bibitem[\protect\citeauthoryear{Rodeghero, Liu, McBurney, and
  McMillan}{Rodeghero et~al\mbox{.}}{2015}]%
        {rodeghero2015eye}
\bibfield{author}{\bibinfo{person}{Paige Rodeghero}, \bibinfo{person}{Cheng
  Liu}, \bibinfo{person}{Paul~W McBurney}, {and} \bibinfo{person}{Collin
  McMillan}.} \bibinfo{year}{2015}\natexlab{}.
\newblock \showarticletitle{An eye-tracking study of java programmers and
  application to source code summarization}.
\newblock \bibinfo{journal}{\emph{IEEE Transactions on Software Engineering}}
  \bibinfo{volume}{41}, \bibinfo{number}{11} (\bibinfo{year}{2015}),
  \bibinfo{pages}{1038--1054}.
\newblock


\bibitem[\protect\citeauthoryear{Roehm, Tiarks, Koschke, and Maalej}{Roehm
  et~al\mbox{.}}{2012}]%
        {Roehm:2012:PDC:2337223.2337254}
\bibfield{author}{\bibinfo{person}{Tobias Roehm}, \bibinfo{person}{Rebecca
  Tiarks}, \bibinfo{person}{Rainer Koschke}, {and} \bibinfo{person}{Walid
  Maalej}.} \bibinfo{year}{2012}\natexlab{}.
\newblock \showarticletitle{How do professional developers comprehend
  software?}. In \bibinfo{booktitle}{\emph{Proceedings of the 2012
  International Conference on Software Engineering}}
  \emph{(\bibinfo{series}{ICSE 2012})}. \bibinfo{publisher}{IEEE Press},
  \bibinfo{address}{Piscataway, NJ, USA}, \bibinfo{pages}{255--265}.
\newblock
\showISBNx{978-1-4673-1067-3}
\urldef\tempurl%
\url{http://dl.acm.org/citation.cfm?id=2337223.2337254}
\showURL{%
\tempurl}


\bibitem[\protect\citeauthoryear{Roscher, Bohn, Duarte, and Garcke}{Roscher
  et~al\mbox{.}}{2019}]%
        {roscher2019explainable}
\bibfield{author}{\bibinfo{person}{Ribana Roscher}, \bibinfo{person}{Bastian
  Bohn}, \bibinfo{person}{Marco~F. Duarte}, {and} \bibinfo{person}{Jochen
  Garcke}.} \bibinfo{year}{2019}\natexlab{}.
\newblock \bibinfo{title}{Explainable Machine Learning for Scientific Insights
  and Discoveries}.
\newblock
\newblock
\showeprint[arxiv]{cs.LG/1905.08883}


\bibitem[\protect\citeauthoryear{Samek, Wiegand, and M{\"u}ller}{Samek
  et~al\mbox{.}}{2017}]%
        {samek2017explainable}
\bibfield{author}{\bibinfo{person}{Wojciech Samek}, \bibinfo{person}{Thomas
  Wiegand}, {and} \bibinfo{person}{Klaus-Robert M{\"u}ller}.}
  \bibinfo{year}{2017}\natexlab{}.
\newblock \showarticletitle{Explainable artificial intelligence: Understanding,
  visualizing and interpreting deep learning models}.
\newblock \bibinfo{journal}{\emph{arXiv preprint arXiv:1708.08296}}
  (\bibinfo{year}{2017}).
\newblock


\bibitem[\protect\citeauthoryear{Shi, Zhong, Xie, and Li}{Shi
  et~al\mbox{.}}{2011}]%
        {Shi:2011:ESE:1987434.1987473}
\bibfield{author}{\bibinfo{person}{Lin Shi}, \bibinfo{person}{Hao Zhong},
  \bibinfo{person}{Tao Xie}, {and} \bibinfo{person}{Mingshu Li}.}
  \bibinfo{year}{2011}\natexlab{}.
\newblock \showarticletitle{An empirical study on evolution of API
  documentation}. In \bibinfo{booktitle}{\emph{Proceedings of the 14th
  international conference on Fundamental approaches to software engineering:
  part of the joint European conferences on theory and practice of software}}
  \emph{(\bibinfo{series}{FASE'11/ETAPS'11})}.
  \bibinfo{publisher}{Springer-Verlag}, \bibinfo{address}{Berlin, Heidelberg},
  \bibinfo{pages}{416--431}.
\newblock
\showISBNx{978-3-642-19810-6}
\urldef\tempurl%
\url{http://dl.acm.org/citation.cfm?id=1987434.1987473}
\showURL{%
\tempurl}


\bibitem[\protect\citeauthoryear{Song, Sun, Wang, and Yan}{Song
  et~al\mbox{.}}{2019}]%
        {song2019survey}
\bibfield{author}{\bibinfo{person}{Xiaotao Song}, \bibinfo{person}{Hailong
  Sun}, \bibinfo{person}{Xu Wang}, {and} \bibinfo{person}{Jiafei Yan}.}
  \bibinfo{year}{2019}\natexlab{}.
\newblock \showarticletitle{A Survey of Automatic Generation of Source Code
  Comments: Algorithms and Techniques}.
\newblock \bibinfo{journal}{\emph{IEEE Access}} (\bibinfo{year}{2019}).
\newblock


\bibitem[\protect\citeauthoryear{Sridhara, Hill, Muppaneni, Pollock, and
  Vijay-Shanker}{Sridhara et~al\mbox{.}}{2010}]%
        {sridhara2010towards}
\bibfield{author}{\bibinfo{person}{Giriprasad Sridhara}, \bibinfo{person}{Emily
  Hill}, \bibinfo{person}{Divya Muppaneni}, \bibinfo{person}{Lori Pollock},
  {and} \bibinfo{person}{K Vijay-Shanker}.} \bibinfo{year}{2010}\natexlab{}.
\newblock \showarticletitle{Towards automatically generating summary comments
  for java methods}. In \bibinfo{booktitle}{\emph{Proceedings of the IEEE/ACM
  international conference on Automated software engineering}}. ACM,
  \bibinfo{pages}{43--52}.
\newblock


\bibitem[\protect\citeauthoryear{Sridhara, Pollock, and Vijay-Shanker}{Sridhara
  et~al\mbox{.}}{2011}]%
        {sridhara2011automatically}
\bibfield{author}{\bibinfo{person}{Giriprasad Sridhara}, \bibinfo{person}{Lori
  Pollock}, {and} \bibinfo{person}{K Vijay-Shanker}.}
  \bibinfo{year}{2011}\natexlab{}.
\newblock \showarticletitle{Automatically detecting and describing high level
  actions within methods}. In \bibinfo{booktitle}{\emph{Proceedings of the 33rd
  International Conference on Software Engineering}}. ACM,
  \bibinfo{pages}{101--110}.
\newblock


\bibitem[\protect\citeauthoryear{Sutskever, Martens, and Hinton}{Sutskever
  et~al\mbox{.}}{2011}]%
        {sutskever2011generating}
\bibfield{author}{\bibinfo{person}{Ilya Sutskever}, \bibinfo{person}{James
  Martens}, {and} \bibinfo{person}{Geoffrey~E Hinton}.}
  \bibinfo{year}{2011}\natexlab{}.
\newblock \showarticletitle{Generating text with recurrent neural networks}. In
  \bibinfo{booktitle}{\emph{Proceedings of the 28th International Conference on
  Machine Learning (ICML-11)}}. \bibinfo{pages}{1017--1024}.
\newblock


\bibitem[\protect\citeauthoryear{Sutskever, Vinyals, and Le}{Sutskever
  et~al\mbox{.}}{2014}]%
        {sutskever2014sequence}
\bibfield{author}{\bibinfo{person}{Ilya Sutskever}, \bibinfo{person}{Oriol
  Vinyals}, {and} \bibinfo{person}{Quoc~V Le}.}
  \bibinfo{year}{2014}\natexlab{}.
\newblock \showarticletitle{Sequence to sequence learning with neural
  networks}. In \bibinfo{booktitle}{\emph{Advances in neural information
  processing systems}}. \bibinfo{pages}{3104--3112}.
\newblock


\bibitem[\protect\citeauthoryear{Von~Mayrhauser and Vans}{Von~Mayrhauser and
  Vans}{1995}]%
        {von1995program}
\bibfield{author}{\bibinfo{person}{Anneliese Von~Mayrhauser} {and}
  \bibinfo{person}{A~Marie Vans}.} \bibinfo{year}{1995}\natexlab{}.
\newblock \showarticletitle{Program comprehension during software maintenance
  and evolution}.
\newblock \bibinfo{journal}{\emph{Computer}} \bibinfo{number}{8}
  (\bibinfo{year}{1995}), \bibinfo{pages}{44--55}.
\newblock


\bibitem[\protect\citeauthoryear{von Rueden, Mayer, Garcke, Bauckhage, and
  Schuecker}{von Rueden et~al\mbox{.}}{2019}]%
        {rueden2019informed}
\bibfield{author}{\bibinfo{person}{Laura von Rueden},
  \bibinfo{person}{Sebastian Mayer}, \bibinfo{person}{Jochen Garcke},
  \bibinfo{person}{Christian Bauckhage}, {and} \bibinfo{person}{Jannis
  Schuecker}.} \bibinfo{year}{2019}\natexlab{}.
\newblock \bibinfo{title}{Informed Machine Learning - Towards a Taxonomy of
  Explicit Integration of Knowledge into Machine Learning}.
\newblock
\newblock
\showeprint[arxiv]{stat.ML/1903.12394}


\bibitem[\protect\citeauthoryear{Wan, Zhao, Yang, Xu, Ying, Wu, and Yu}{Wan
  et~al\mbox{.}}{2018}]%
        {wan2018reinforcement}
\bibfield{author}{\bibinfo{person}{Yao Wan}, \bibinfo{person}{Zhou Zhao},
  \bibinfo{person}{Min Yang}, \bibinfo{person}{Guandong Xu},
  \bibinfo{person}{Haochao Ying}, \bibinfo{person}{Jian Wu}, {and}
  \bibinfo{person}{Philip~S. Yu}.} \bibinfo{year}{2018}\natexlab{}.
\newblock \showarticletitle{Improving Automatic Source Code Summarization via
  Deep Reinforcement Learning}. In \bibinfo{booktitle}{\emph{Proceedings of the
  33rd ACM/IEEE International Conference on Automated Software Engineering}}
  \emph{(\bibinfo{series}{ASE 2018})}. \bibinfo{publisher}{Association for
  Computing Machinery}, \bibinfo{address}{New York, NY, USA},
  \bibinfo{pages}{397–407}.
\newblock
\showISBNx{9781450359375}
\urldef\tempurl%
\url{https://doi.org/10.1145/3238147.3238206}
\showDOI{\tempurl}


\bibitem[\protect\citeauthoryear{Wu, Pan, Chen, Long, Zhang, and Yu}{Wu
  et~al\mbox{.}}{2019}]%
        {wu2019gnnsurvey}
\bibfield{author}{\bibinfo{person}{Zonghan Wu}, \bibinfo{person}{Shirui Pan},
  \bibinfo{person}{Fengwen Chen}, \bibinfo{person}{Guodong Long},
  \bibinfo{person}{Chengqi Zhang}, {and} \bibinfo{person}{Philip~S. Yu}.}
  \bibinfo{year}{2019}\natexlab{}.
\newblock \showarticletitle{A Comprehensive Survey on Graph Neural Networks}.
\newblock \bibinfo{journal}{\emph{CoRR}}  \bibinfo{volume}{abs/1901.00596}
  (\bibinfo{year}{2019}).
\newblock
\showeprint[arxiv]{1901.00596}
\urldef\tempurl%
\url{http://arxiv.org/abs/1901.00596}
\showURL{%
\tempurl}


\bibitem[\protect\citeauthoryear{Xu, Wu, Wang, Feng, Witbrock, and Sheinin}{Xu
  et~al\mbox{.}}{2018a}]%
        {xu2018graph2seq}
\bibfield{author}{\bibinfo{person}{Kun Xu}, \bibinfo{person}{Lingfei Wu},
  \bibinfo{person}{Zhiguo Wang}, \bibinfo{person}{Yansong Feng},
  \bibinfo{person}{Michael Witbrock}, {and} \bibinfo{person}{Vadim Sheinin}.}
  \bibinfo{year}{2018}\natexlab{a}.
\newblock \showarticletitle{Graph2seq: Graph to sequence learning with
  attention-based neural networks}.
\newblock \bibinfo{journal}{\emph{arXiv preprint arXiv:1804.00823}}
  (\bibinfo{year}{2018}).
\newblock


\bibitem[\protect\citeauthoryear{Xu, Wu, Wang, Yu, Chen, and Sheinin}{Xu
  et~al\mbox{.}}{2018b}]%
        {xu2018exploiting}
\bibfield{author}{\bibinfo{person}{Kun Xu}, \bibinfo{person}{Lingfei Wu},
  \bibinfo{person}{Zhiguo Wang}, \bibinfo{person}{Mo Yu},
  \bibinfo{person}{Liwei Chen}, {and} \bibinfo{person}{Vadim Sheinin}.}
  \bibinfo{year}{2018}\natexlab{b}.
\newblock \showarticletitle{Exploiting rich syntactic information for semantic
  parsing with graph-to-sequence model}.
\newblock \bibinfo{journal}{\emph{Conference on Empirical Methods in Natural
  Language Processing}} (\bibinfo{year}{2018}).
\newblock


\end{thebibliography}

\end{document}